\documentclass[12pt]{article}

\newcommand{\be}{\begin{equation}}
\newcommand{\ee}{\end{equation}}
\newcommand{\ba}{\begin{eqnarray}}
\newcommand{\ea}{\end{eqnarray}}
\newcommand{\barr}{\begin{array}}
\newcommand{\earr}{\end{array}}
\def\X {{{\cal X}}}
\def\Y {{{\cal Y}}}
\def\Z {{{\cal Z}}}

\newcommand{\x}{\tilde X}
\newcommand{\y}{\tilde Y}

\begin{document}
\title{Fully Self-consistent RPA description of the many level pairing model}
\author{Jorge G. Hirsch\,$^1$, Alejandro Mariano\,$^2$, Jorge Dukelsky\,$^3$,\\
and Peter Schuck\,$^4$ \\
{\small \it $^1$ Instituto de Ciencias Nucleares, Universidad Nacional}\\
{\small \it Aut\'onoma de M\'exico, Apdo. Postal 70-543 M\'{e}xico 04510 D.F.}\\
{\small \it $^2$ Departamento de F\'{\i}sica, Universidad Nacional de La Plata,}\\ 
{\small \it C.C. 67, La Plata (1900), Argentina }\\
{\small \it $^3$ Instituto de Estructura de la Materia, Consejo Superior de}\\
{\small \it Investigaciones Cient\'\i ficas, Serrano 123, 28006 Madrid, Spain}\\
{\small \it $^4$ Institut de Physique  Nucl\'{e}aire, Universit\'e Paris-Sud,}\\
{\small \it F-91406 Orsay Cedex, France}\\
}
\maketitle

\begin{abstract}
The Self-Consistent RPA (SCRPA) equations in the particle-particle
channel are solved without any approximation for the picket fence model.
The results are in excellent agreement with the exact solutions found
with the Richardson method. Particularly interesting features are that
screening corrections reverse the sign of the interaction and that SCRPA
yields the exact energies in the case of two levels with two particles.
\end{abstract}



This manuscript has 37 pages, 1 figure and 17 tables.

\newpage

Running head: {\em Fully SCRPA description of the ...}

\bigskip \bigskip

\begin{tabular}{ll}
Correspondent author: &Dr. Jorge G. Hirsch\\
	&Instituto de Ciencias Nucleares, \\
	&Universidad Nacional Aut\'onoma de M\'exico,\\
	&Apdo. Postal 70-543 M\'{e}xico 04510 D.F.\\
	&phone: 52 56224672,  fax: 52 56162233 \\
	&e-mail: hirsch@nuclecu.unam.mx
\end{tabular}

\newpage

\section{Introduction}

The approach to the many body problem is often divided into the evaluation of a hierarchy of
correlation functions. First comes the one body or mean field theory. Second
the two body correlation functions, and so on. There is a large
consensus on the mean field or Hartree-Fock level so  that
practically in any field of theoretical physics one basically understands
the same by this approach in spite of the fact that there may be differences
in detail. However, already on the next level of complication, $i.e.$
considering two body correlations, the strategies in various fields start to
diverge strongly. Let us mention first a few of them: Brueckner-Hartree-Fock 
\cite{Day67}, Jastrow ansatz for correlated wave functions, the hypernetted
chain method, correlated basis functions \cite{Fan84}, coupled cluster theory 
\cite{Kum78}, Gutzwiller ansatz for correlated ground states \cite{Blai86}, and
quite a few more are examples of different strategies to tackle the problem
of strongly correlated quantum systems. There seems no common consensus
about which of the strategies is best, all seem to be more or less tailored
for specific problems. In this situation it seems legitimate to explore further
routes which have been not sufficiently exploited in the past. It is the
latter case which we are investigating since a couple of years, further
developing the so-called Self-Consistent RPA (SCRPA) theory which has been
initiated mostly in nuclear physics with the pioneering work of Hara \cite{Har64},
and further developed by Rowe \cite{Row68}. 
Independently and using different techniques the same
approach has been developed in plasma physics where it is called cluster
Hartree-Fock theory \cite{Rop95}. Very early quite similar ideas have also
been put forward by the Japanese school of Marumori et al \cite{Mar64}. The
recent theoretical developments along this line and the various successes in
applications to model cases \cite{Schu71,Schu73,Duk90,Duk96,Duk98a,Duk98b,Duk99,Duk00} 
make us firmly believe that there exists a great
potentiality in this approach which is very general and can be easily
adapted to practically any kind of many body problem where two body
correlations or even higher correlations are of importance.

The SCRPA scheme has indeed, on the level of the two body correlations or
fluctuations characterized by a correlated pair of Fermion operators
like $Q_{\nu }^{\dagger }=\sum_{kk^{\prime }}X_{kk^{\prime }}^{\nu
}a_{k}^{\dagger }a_{k^{\prime }}^{\dagger }$ or $\sum_{kk^{\prime
}}X_{kk^{\prime }}^{\nu }a_{k}^{\dagger }a_{k^{\prime }}$, aspects quite
analogous to ordinary Hartree-Fock theory characterized by the single
operator $q_{\alpha }^{\dagger }=\sum_{k}\varphi _{k}^{\alpha
}a_{k}^{\dagger }$ . As we will show in the main text, standard
Hartree-Fock theory and SCRPA or Cluster Hartree-Fock theory can be derived
from a common variational principle leading to the equations

\begin{equation}
\left\langle \left\{ \delta q , \left[ H,q_{\alpha }^{\dagger }\right] \right\}
\right\rangle =\varepsilon _{\alpha }\left\langle \left\{ \delta q,q_{\alpha
}^{\dagger }\right\} \right\rangle ~,  \label{hf1}
\end{equation}

\begin{equation}
\left\langle \left[ \delta Q , \left[ H,Q_{\nu }^{\dagger }\right] \right]
\right\rangle =E_{\nu }\left\langle \left[ \delta Q,Q_{\nu }^{\dagger
}\right] \right\rangle   ~,\label{rpa1}
\end{equation}
where $\left\{ ,\right\} $ is the anticommutator and $\left[ ,\right] \,$the
commutator, and $\left\langle \cdots \right\rangle $ stands for the mean
value with respect to vacua defined by $q_{\alpha }\left| 0\right\rangle
= 0$ or $Q_{\nu }\left| 0\right\rangle = 0$. 
Equations (\ref{hf1}) and (\ref{rpa1}) are not only
formally very similar. Also their physical content is quite analogous.
Whereas (\ref{hf1}) describes the motion of one particle in the average
field created by all the others, the second equation (\ref{rpa1}) describes
the motion of a quantal fluctuation (a correlated Fermion pair) in the average 
field created by all the other fluctuations (pairs). 
A nucleus, for example, can be considered in some approximation as a gas of 
independent zero point vibrations (elementary excitations) as well as a gas 
of independent particles. 
Both gases create their own mean field and in general they will be coupled.

As with any theory for correlation functions also SCRPA is very difficult
and demanding with respect to numerical implementation in the general case.
However the recent solution of the many level pairing model is sufficiently
complex to state that with respect to a brute force diagonalization, the
SCRPA gives a tremendous simplification of the solution of the many body
problem with very good accuracy which, at least for the case considered, does
not fail in any qualitative respects. It is our intention in this work to
take up again the SCRPA solution of the many level pairing model of Ref. 
\cite{Duk99} in more detail. In first place we circumvent an approximation used
in \cite{Duk99} so that in this work the SCRPA scheme is carried through in
full without any approximation. On the basis of this achievement we will
then be able to discuss the advantages and eventual shortcomings of the
method. We also will be able to assess the deficiency of the approximation
we adopted in \cite{Duk99} as well as other more severe simplifying reductions
of the full SCRPA scheme such as the renormalized RPA (r-RPA) which, because
of its relatively simple numerical structure, has become quite popular in
the recent past \cite{Cat94,Toi95,Cat96}. 

In detail the paper is organized as follows: in Section 2 the picket-fence model
is introduced, in Section 3 it is reformulated to make explicit its particle-hole
symmetry, in Section 4 the SCRPA formalism is presented, in Section 5 the occupation 
numbers and correlation functions needed to close the equations are constructed, in 
Section 6 the approximations used to obtain simpler RPA formalisms are discussed, 
in Section 7 sum rules related with the conservation of the particle number are
presented. The explicit construction and the problems associated with the SCRPA ground 
state are discussed in Section 8, numerical results are shown in Section 9, and the 
conclusions in Section 10.
Some useful mathematical results are given in the Appendix.

\section{The Picket-Fence Model}

The picket fence or multilevel pairing model has been introduced by Richardson in 1966 
\cite{Rich66}
to describe deformed nuclei. The model has the great advantage of being non-trivial 
(in the sense that it can not straightforwardly be diagonalised for an arbitrary number 
of levels) and still being exactly solvable (by some kind of Bethe ansatz).
However, in nuclear physics, aside from some rare considerations \cite{Ban70}, 
the model has not been 
exploited very much probably because it has been judged too crude for the description of 
real nuclei. Still the model contains very interesting physics and it has recently been 
revived in the context of ultra-small superconducting metallic grains \cite{Bra98}.
One of the most interesting aspects of the model is that the exact solution reveals a 
transition between the superfluid (or superconducting) regime and the normal state which 
is completely smooth, $i.e.$ no sign of any abrupt phase transition from one state to the 
other can be detected as a function of the system parameters \cite{Duk99}.

The original Hamiltonian of the model is given by

\be
H=\sum_{i=1}^{\Omega }\left( \varepsilon _{i}-\lambda \right)
N_{i}-G\sum_{i,j=1}^{\Omega }P_{i}^{\dagger }P_{j} ~,\label{ham}
\ee
with
\begin{equation}
N_i = c^\dagger_i c_i + c^\dagger_{-i} c_{-i}, ~~
P^\dagger_i = c^\dagger_i c^\dagger_{-i}  ~,\label{ni}
\end{equation}
where $c^\dagger_i$  creates a particle in the i-th level with $j = {\frac
1 2}$ and $ m = {\frac 1 2}$ and $c^\dagger_{-i}$ with $ m = -{\frac 1
2}$. $\Omega $ is the total number of levels, $G$ is the pairing
interaction strength and $\varepsilon _{i}=i\ \varepsilon$. 
The chemical potential $\lambda $ will be
defined such that the Hamiltonian preserves particle-hole symmetry.
Each level has degeneracy two and we will assume that the system is half
filled with number of pairs $N=\Omega /2$ .

The particle ($p$) and hole ($h$) states are defined by
\begin{equation}
N_h |HF\rangle = 2, ~~~N_p |HF\rangle = 0 ~,
\end{equation}
where $|HF\rangle$ is the ground state of the Hamiltonian (\ref{ham}) with
G=0.
The particle states $p$ correspond to $\varepsilon _{p}>\lambda $
and the holes $h$ to $\varepsilon _{p}<\lambda $.

In this case, with no partial occupations allowed, the following relation
is fulfilled:

\begin{equation}
P^\dagger_i P_i + P_i P^\dagger_i = 1 ~,
\end{equation}
which implies
\begin{equation}
N_i = 2 P^\dagger_i P_i  ~.\label{np}
\end{equation}

\section{The particle-hole symmetry}

The commutation relations between the operators defined in (2) are

\be
\left[ P_{i},P_{j}^{\dagger }\right] =\delta _{ij}\left( 1-N_{i}\right)~,
\hbox{~~~~~}
\left[ N_{i},P_{j}^{\dagger }\right] =2\delta _{ij}P_{j}^{\dagger } ~,
\hbox{~~~~~}
\left[ N_{i},P_{j}\right] =-2\delta _{ij}P_{j} ~.
\ee

To make explicit the particle-hole symmetry we will make the following
particle-hole conjugation

\be
c_{p}=d_{p}\quad ,\quad c_{\overline{p}}=d_{\overline{p}}
\quad ,\quad \quad \quad
c_{h}=d_{\overline{h}}^{\dagger }\quad ,\quad c_{\overline{h}%
}=-d_{h}^{\dagger } .
\ee
The new operators $M$, $Q$
and $Q^{\dagger }$ in terms of $d^{\dagger }\left( d\right) $ are
\be
N_{h}=2-M_{h},\quad N_{p}=M_{p} ,\quad \quad
P_{h}^{\dagger }=-Q_{h}\quad ,\quad P_{p}^{\dagger }=Q_{p}^{\dagger }  ~.\label{mp}
\ee
Their commutation relations are

\ba
\left[ Q_{p},Q_{p'}^{\dagger }\right] =\delta _{pp'}\left( 1-M_{p}\right)~,
\quad
\left[ M_{p},Q_{p'}^{\dagger }\right] =2\delta _{pp'}Q_{p}^{\dagger }~,\quad
\quad
\left[ M_{p},Q_{p'}\right] =-2\delta _{pp'}Q_{p}~,\quad \nonumber \\
\left[ Q_{h},Q_{h'}^{\dagger }\right] =\delta _{hh'}\left( 1-M_{h}\right)~,
\quad
\left[ M_{h},Q_{h'}^{\dagger }\right] =2\delta _{hh'}Q_{h}^{\dagger }~,\quad
\quad
\left[ M_{h},Q_{h'}\right] =-2\delta _{hh'}Q_{h} ~.
\ea

The single particle energies are $\varepsilon _{p}=\varepsilon (N+p)$ and
$\varepsilon_{h}= \varepsilon (N-h+1)$, with $p,h=1,\ldots ,N$ , and
$N=\frac{1}{2}\Omega $. Particles (p) and holes (h) are numbered starting
from the level closest to the Fermi level.  
We use the chemical potential to restore the particle-hole symmetry:

\be
\lambda =\varepsilon \left( N+\frac{1}{2}\right) -\frac{G}{2}~.
\ee

With this definition the Hamiltonian reduces to
  
\ba
H=-\varepsilon N^{2}+\sum_{p=h=1}^{N}\left[ \varepsilon \left(
p-\frac{1}{2}%
\right) +\frac{G}{2}\right] \left( M_{p}+M_{h}\right) 
\nonumber \\
-G\sum_{pp^{\prime}}Q_{p}^{\dagger }Q_{p^{\prime }}
-G\sum_{hh^{\prime }}Q_{h}^{\dagger }Q_{h^{\prime }}
+G\sum_{ph}\left( Q_{p}^{\dagger }Q_{h}^{\dagger }+Q_{p}Q_{h}\right) ~.
\ea

In this form the complete symmetry between particle and hole states becomes evident.This will
greatly facilitate later the formal and numerical aspects of our theory.

\section{The RPA formalism}

The basic ingredients of the SCRPA approach in the particle-particle channel are the two 
particle addition operator
  
\be
A_{\mu }^{\dagger }=\sum_{p}X_{p}^{\mu }\ \overline{Q}_{p}^{\dagger
}-\sum_{h}Y_{h}^{\mu }\ \overline{Q}_{h} ~, \label{A}
\ee
and the removal operator

\be
R_{\lambda }^{\dagger }=-\sum_{p}Y_{p}^{\lambda }\overline{Q}
_{p}+\sum_{h}X_{h}^{\lambda }\overline{Q}_{h}^\dagger  ~,\label{R}
\ee
where $\overline{Q}_{p} = Q_p / \sqrt{1 - \langle M_p \rangle}$ and
$\overline{Q}_{h} = Q_h / \sqrt{1 - \langle M_h \rangle}$. 

Following Baranger in his derivation for the single particle mean field equations 
\cite{Bar60}, we define the following mean excitation energy

\ba
\Omega_{\mu} =  &\left\{ \sum_{\alpha (N+2)}(E^{\alpha}_{N+2}-E^{0}_{N+2}) 
|\langle \alpha|A^\dagger_{\mu}|0 \rangle |^2 
\right. \nonumber \\
&+ \sum_{\beta (N-2)}(E^{\beta}_{N-2}-E^{0}_{N-2})| \langle \beta |A_{\mu}|0 \rangle |^2 
\nonumber \\
& \left. 
+ 2 \mu^{(+)} \sum_{\alpha} | \langle \alpha |A^\dagger_{\mu} |0 \rangle |^2 
- 2 \mu^{(-)} \sum_{\beta} | \langle \beta |A_{\mu} |0 \rangle |^2] \right\} / 
\label{omega}\\
& \left\{ \sum_{\alpha (N+2)} | \langle \alpha |A^\dagger_{\mu} |0 \rangle |^2 
- \sum_{\beta (N-2)} | \langle \beta |A_{\mu} |0 \rangle |^2 \right\}  ~,
\nonumber
\ea
where $2 \mu^{(\pm)}=(\pm) 1/2 (E^{0}_{N+/-2} - E^{0}_N)$ are the chemical potentials and 
$E^{\alpha,\beta}_N $ are the in principle exact eigen values of $H$ of Eq.(1) and 
$|\alpha \rangle ,| \beta \rangle , | 0 \rangle $ are the corresponding exact eigenstates.

Expression (\ref{omega}) can be considered as an average excitation energy taking into 
account the 
spectra of both the $N+2$ and $N-2$ systems. In fact $\Omega_{\mu}$ represents the energy 
weighted sum rule over both spectra divided by the non-energy weighted sum rule.

Indeed expression (\ref{omega}) can also be written as 
\be
\Omega_{\mu}=\frac{ \langle 0|[A_{\mu},[H,A^\dagger_{\mu}]] | 0 \rangle}
{\langle 0|[A_{\mu},A^ \dagger_{\mu}] |0 \rangle}  ~,
\ee
which makes the analogy with the usual energy weighted sum rule in the particle-hole channel 
quite obvious. The terms in (\ref{omega}) involving the chemical potentials $\mu^{(\pm)}$ are
needed in order to give the correct origin of the energy spectra.
This is best seen in supposing that $\mu^{(+)}= \mu^{(-)}$ which leads to 
\be
\Omega_{\mu}-2 \mu = \frac{\sum_{\alpha}(E^{\alpha}_{N+2}-E^{0}_{N+2})
|\langle\alpha |A^\dagger_{\mu} |0 \rangle |^2 
+ \sum_{\beta}(E^{\beta}_{N-2}-E^{0}_{N-2}) 
| \langle \beta |A_{\mu} |0 \rangle |^2}{\langle 0 |[A_{\mu},A^\dagger_{\mu}] | 0 \rangle} .
\ee
It means that the energy weighting starts from two times the chemical potential,
as it should be.

The first step is now to minimize the average two particle energy (\ref{omega}) 
with respect to the amplitudes $X, Y$. This leads straightforwardly to the following 
set of equations:
\be
\left( \barr{cc} A & B \\ -B & C \earr \right)
\left( \barr{c} X \\ Y \earr \right)
= E
\left( \barr{c} X \\ Y \earr \right) ~,\label{rpa}
\ee
where
\ba
A_{pp'} = 
\langle 0|[\overline{Q}_{p},[H,\overline{Q}_{p'}^\dagger]] | 0 \rangle = \nonumber \\
\delta_{pp'} \left\{ 2 \left(\epsilon (p - {\frac 1 2}) +
{\frac G 2} \right)
+ 2 {\frac {G} {1 - \langle M_p \rangle}} 
\langle (\sum_{p_1} Q^\dagger_{p_1} - \sum_{h_1} Q_{h_1} ) Q_p \rangle
\right\} \nonumber\\
- G {\frac {\langle (1 - M_p) (1 - M_{p'}) \rangle}
   {\sqrt{(1 - \langle M_p \rangle) (1 -\langle  M_{p'}\rangle)} } } ~,
\nonumber \\
B_{ph} = \langle 0|[\overline{Q}_{p},[H,\overline{Q}_{h}^\dagger]] | 0 \rangle = 
 G {\frac {\langle (1 - M_p) (1 - M_{h}) \rangle}
   {\sqrt{(1 - \langle M_p \rangle) (1 -\langle  M_{h}\rangle)} } } ~,
\label{scrpa}\\
C_{hh'} = \langle 0|[\overline{Q}_{h},[H,\overline{Q}_{h'}^\dagger]] | 0 \rangle = 
\nonumber \\
\delta_{hh'} \left\{ -2 \left(\epsilon (h - {\frac 1 2}) + 
{\frac G 2} \right)
- 2 {\frac {G} {1 - \langle M_h \rangle}}
\langle Q^\dagger_h (-\sum_{p_1} Q^\dagger_{p_1} + \sum_{h_1} Q_{h_1} ) 
\rangle \right\} \nonumber \\
+ G {\frac {\langle (1 - M_h) (1 - M_{h'}) \rangle}
   {\sqrt{(1 - \langle M_h \rangle) (1 -\langle  M_{h'}\rangle)} } } ~.
\nonumber 
\ea

Due to the particle-hole symmetry, the removal mode satisfy exactly the
same equations, implying that both modes have exactly the same excitation
energies and wave functions. To advance this conclusion we are assuming
the following relations:

\ba
\langle M_{p} \rangle  = \langle M_{h=p} \rangle ,\nonumber \\
\langle Q^\dagger_{p} Q_{p'}\rangle = 
\langle Q^\dagger_{h=p} Q_{h'=p'}\rangle, \quad 
\langle Q_{h} Q_{p}\rangle = 
\langle Q^\dagger_{p} Q^\dagger_{h}\rangle \label{adv} \\
\langle M_{p} M_{p'} \rangle  = \langle M_{h=p} M_{h'=p'} \rangle ,\quad
\langle M_{h} M_{p} \rangle  = \langle M_{h'=p} M_{p'=h} \rangle
 , \nonumber  
\ea
which are shown below to be consistent with the above equations.

It means that

\be
X^{\mu }_{p } = \pm X^{\lambda = \mu}_{h = p}, \quad \quad
Y^{\mu }_{h } = \pm Y^{\lambda = \mu}_{p = h} .\label{x=x}
\ee
The forward and backward amplitudes $X,Y$ fulfill the normalization
conditions
\ba
\sum_{p}X_{p}^{\mu }X_{p}^{\mu ^{\prime }}-\sum_{h}Y_{h}^{\mu }Y_{h}^{\mu
^{\prime }}=\delta _{\mu \mu ^{\prime }} ~,\nonumber \\
\sum_{h}X_{h}^{\lambda }X_{h}^{\lambda ^{\prime }}-\sum_{p}Y_{p}^{\lambda
}Y_{p}^{\lambda ^{\prime }}=\delta _{\lambda \lambda ^{\prime }} ~,\\
\sum_{p}X_{p}^{\mu} Y_{p}^{\lambda }-\sum_{h}X_{h}^{\lambda} Y_{h}^{\mu} 
= 0  ~,\nonumber
\ea
and the closure relations
  \ba
\sum_{\mu }X_{p}^{\mu }X_{p^{\prime }}^{\mu }-\sum_{\lambda
}Y_{p}^{\lambda
}Y_{p^{\prime }}^{\lambda }=\delta _{pp^{\prime }} ~,\nonumber \\
\sum_{\lambda }X_{h}^{\lambda }X_{h^{\prime }}^{\lambda }-\sum_{\mu
}Y_{h}^{\mu }Y_{h^{\prime }}^{\mu }=\delta _{hh^{\prime }}  ~,\\
\sum_{\lambda }X_{h}^{\lambda }Y_{p}^{\lambda }-\sum_{\mu }X_{p}^{\mu
}Y_{h}^{\mu }=0 ~.\nonumber
\ea

The expectation values of the commutators are

\ba
\left\langle \left[ R_{\lambda },R_{\lambda ^{\prime }}^{\dagger }\right]
\right\rangle =\delta _{\lambda \lambda ^{\prime }} ~,\nonumber \\
\left\langle \left[ A_{\mu },A_{\mu ^{\prime }}^{\dagger }\right]
\right\rangle =\delta _{\mu \mu ^{\prime }} ~,\\
\left\langle \left[ A_{\mu }^{\dagger },R_{\lambda ^{\prime }}^{\dagger
}\right] \right\rangle =\left\langle \left[ A_{\mu },R_{\lambda ^{\prime
}}\right] \right\rangle =0 ~,\nonumber \\
\left[ R_{\lambda },A_{\mu }^{\dagger }\right] = 0 ~.
\ea

With the help of these equations we can now invert (\ref{A},\ref{R}) which yields
  
\ba
Q_{p}^{\dagger }=\sqrt{1-\left\langle M_{p}\right\rangle }\left[ \sum_{\mu
}X_{p}^{\mu }A_{\mu }^{\dagger }+\sum_{\lambda }Y_{p}^{\lambda }R_{\lambda
}\right] ~,\nonumber \\
Q_{h} =\sqrt{1 - \left\langle M_{h}\right\rangle }\left[
\sum_{\lambda }X_{h}^{\lambda }R_{\lambda }+\sum_{\mu }Y_{h}^{\mu }A_{\mu
}^{\dagger }\right] ~.\label{qpqh}
\ea

\smallskip An important point to recognize at this step is the fact that the above 
SCRPA equations can also be derived using the equation of motion method advocated 
for example by Rowe \cite{Row68}. A basic ingredient to this latter method is 
that one assumes the existence of a vacuum state $| 0 \rangle \equiv |SCRPA \rangle$ such that

\be
A_{\mu }\left| SCRPA\right\rangle = R_{\lambda }\left| SCRPA\right\rangle = 0  .
\label{vac-rpa}
\ee

Therefore Eq. (\ref{vac-rpa}) must be considered as an inherent additional relation 
belonging to the whole SCRPA scheme. The expectation values in (\ref{scrpa}) shall then be 
evaluated with this ground state and, as we will show in a moment, this will permit 
to entirely close the system of equations.

\section{Closing the system of equations}

The aim is now to express the expectation values which figure in the RPA-matrix (\ref{scrpa})
entirely and without approximation by the amplitudes $X,Y$. Using (\ref{qpqh}, \ref{vac-rpa}) 
this can easily be achieved for most of the expectation values. Indeed one directly verifies:

\ba
\langle Q^\dagger_{p} Q_{p'}\rangle = 
\sqrt{(1-\langle M_p \rangle )(1-\langle M_{p'} \rangle )}
\sum_\lambda Y^\lambda_p Y^\lambda_{p'}, \nonumber \\
\langle Q^\dagger_{h} Q_{h'}\rangle =
\sqrt{(1-\langle M_h \rangle )(1-\langle M_{h'} \rangle )}
\sum_\mu Y^\mu_h Y^\mu_{h'}, \label{qq}\\
\langle Q_{h} Q_{p}\rangle = \langle Q^\dagger_{p} Q^\dagger_{h}\rangle =
\sqrt{(1-\langle M_h \rangle )(1-\langle M_{p} \rangle )}
\sum_\lambda Y^\lambda_p X^\lambda_{h}  .\nonumber
\ea
Using (\ref{np},\ref{mp}) it is direct to show that

\be
\langle M_p \rangle = 1 - {\frac {1} {1 + 2 \sum\limits_\lambda (
Y^\lambda_p )^2 }} , \quad \quad
\langle M_h \rangle = 1 - {\frac {1} {1 + 2 \sum\limits_\mu (
Y^\mu_h )^2 }} ~, \label{m}
\ee
which together with (\ref{x=x}) implies
\be
\langle M_{p=i} \rangle  = \langle M_{h=i} \rangle \label{m=m} ~,
\ee
reflecting again the particle-hole symmetry.
Using (\ref{x=x}) and (\ref{m=m}) it is also direct to show that
\be
\langle Q^\dagger_{p} Q_{p'}\rangle =
\langle Q^\dagger_{h=p} Q_{h'=p'}\rangle .
\ee
These are the first three of the five equalities advanced in (\ref{adv}).

Knowing these expectation values we can evaluate the SCRPA ground state
energy:
\ba
\langle H \rangle =-\varepsilon N^{2}+\sum_{p=h=1}^{N}\left[ \varepsilon
\left(p-\frac{1}{2}\right) +\frac{G}{2}\right] 
\left( \langle M_{p} \rangle + \langle M_{h} \rangle \right) 
\nonumber \\
-G\sum_{pp^{\prime}} \langle Q_{p}^{\dagger }Q_{p^{\prime }} \rangle 
-G\sum_{hh^{\prime }}\langle Q_{h}^{\dagger }Q_{h^{\prime }} \rangle 
+G\sum_{ph}\left( \langle Q_{p}^{\dagger }Q_{h}^{\dagger }+Q_{p}Q_{h}
\rangle \right) ~.
\ea

The SCRPA correlation energy is
\be
E^{SCRPA}_{corr} = \langle H \rangle +
\varepsilon N^{2} .
\ee
For comparison, the RPA correlation energy is
\be
E^{RPA}_{corr} = -\sum\limits_\mu E_\mu
\sum\limits_p |Y^\mu_p|^2 .
\ee

In order to fully close the set of SCRPA equations we still must express
the correlation functions $\langle M_{i} M_{j} \rangle$ through the RPA
amplitudes. This can also be done exactly, though it is somewhat involved.
For this reason we approximated $\langle M_{i} M_{j} \rangle$ by
$\langle M_{i} \rangle \langle M_{j} \rangle$ in \cite{Duk99}. 
In what follows we now present the full derivation.

 Using (\ref{ni}) and (\ref{np}) it is direct to demonstrate that
\be
N_i N_i = 2 N_i ,
\ee
which implies that
\be
M_p M_p = 2 M_p , \quad M_h M_h = 2 M_h .
\ee

It is also simple to demonstrate that
\be
N_i N_j = 4 \, P^\dagger_i P^\dagger_j P_j P_i \hbox{~~for~~} i\neq j ~,
\ee
which implies the three sets of equations
\ba
M_p M_{p'} =& 4 \, Q^\dagger_p Q^\dagger_{p'} Q_{p'} Q_p \hbox{~~~~~~~~~~~~~~for~~}
p\neq p' ~,\nonumber \\
M_p M_h =&  M_p + M_h - 2 \, Q^\dagger_p Q_h Q^\dagger_h  Q_p 
- 2 \, Q^\dagger_h Q_p Q^\dagger_p  Q_h  ~,\\
M_h M_{h'} =& 4 \, Q^\dagger_h Q^\dagger_{h'} Q_{h'} Q_h \hbox{~~~~~~~~~~~~~~for~~}
h\neq h' ~.\nonumber
\ea

We have then $\Omega$ expectation values which are known:
\be
\langle M_p M_p \rangle = 2 \langle M_p \rangle , ~~
\langle M_h M_h \rangle = 2 \langle M_h \rangle .
\ee

For  $p\neq p', h\neq h'$ the equations are
\ba
\hbox{~~~}\langle M_p M_{p'} \rangle = 
&4 (1- \langle M_{p}\rangle) (1- \langle M_{p'} \rangle)
\sum\limits_{\lambda \lambda'} \sum\limits_{\lambda_1 \lambda_2}
Y^{\lambda}_{p} Y^{\lambda'}_{p} Y^{\lambda_1}_{p'} Y^{\lambda_2}_{p'}
\langle R_\lambda R_{\lambda_1} R^\dagger_{\lambda_2} R^\dagger_{\lambda'}
\rangle ~,\nonumber 
\\
\langle M_p M_{h} \rangle =  & \langle M_p \rangle + \langle M_h \rangle  
\label{mm} \\
&- 2 (1- \langle M_{p} \rangle) (1 - \langle M_{h} \rangle ) 
\sum\limits_{\lambda \lambda'} \sum\limits_{\lambda_1 \lambda_2}   
Y^{\lambda}_{p} Y^{\lambda'}_{p} X^{\lambda_1}_{h} X^{\lambda_2}_{h}
\langle R_\lambda R_{\lambda_1} R^\dagger_{\lambda_2} R^\dagger_{\lambda'}
\rangle \nonumber \\
&- 2 (1- \langle M_{p} \rangle) (1 - \langle M_{h} \rangle )
\sum\limits_{\mu \mu'} \sum\limits_{\mu_1 \mu_2}   
Y^{\mu}_{h} Y^{\mu'}_{h} X^{\mu_1}_{p} X^{\mu_2}_{p}
\langle A_\mu A_{\mu_1} A^\dagger_{\mu_2} A^\dagger_{\mu'}
\rangle ~,\nonumber 
\\
\hbox{~~~}\langle M_h M_{h'} \rangle = 
&4 (1 - \langle M_{h}\rangle ) (1- \langle M_{h'} \rangle )
\sum\limits_{\mu \mu'} \sum\limits_{\mu_1 \mu_2}   
Y^{\mu}_{h} Y^{\mu'}_{h} Y^{\mu_1}_{h'} Y^{\mu_2}_{h'}
\langle A_\mu A_{\mu_1} A^\dagger_{\mu_2} A^\dagger_{\mu'}
\rangle  ~.
\nonumber
\ea

Using the identities given in  the Appendix
we build up the system of equations 

\begin{eqnarray}
&&\hspace{-0.5cm} \sum_{h_1 h_2} \langle M_{h_1} M_{h_2} \rangle
( B^{hh'}_{h_1h_2}- {\frac{\delta_{h h_1} \delta_{h' h_2}}
{4 (1-\langle M_{h_1} \rangle)^2 (1- \langle M_{h_2}\rangle )^2} })
 + \nonumber \\ 
&&\hspace{-0.5cm} \sum_{h_1 p_1} \langle M_{h_1} M_{p_1} \rangle
C^{hh'}_{h_1p_1} + \label{eqb}\\
&&\hspace{-0.5cm} \sum_{p_1 p_2} \langle M_{p_1} M_{p_2} \rangle
D^{hh'}_{p_1p_2}  = F_{hh'}  ~,\label{system} \nonumber
\end{eqnarray}

\begin{eqnarray}
&&\hspace{-0.5cm} \sum_{h_1 h_2} \langle M_{h_1} M_{h_2} \rangle
B^{hp}_{h_1h_2} + \nonumber \\ 
&& \hspace{-0.5cm}\sum_{h_1 p_1} \langle M_{h_1} M_{p_1} \rangle
(C^{hp}_{h_1 p_1} + {\frac {\delta_{h h_1} \delta_{p p_1} }
{2 (1- \langle M_{h_1} \rangle )^2 (1- \langle M_{p_1} \rangle)^2 } } ) +
\label{eqc}\\
&&\hspace{-0.5cm}\sum_{p_1 p_2} \langle M_{p_1} M_{p_2} \rangle 
D^{hp}_{p_1p_2}   = F_{hp} ~,\nonumber 
\end{eqnarray}

\begin{eqnarray}
&& \sum_{h_1 h_2} \langle M_{h_1} M_{h_2} \rangle
B^{pp'}_{h_1h_2} + \nonumber \\ 
&& \sum_{h_1 p_1}\langle M_{h_1} M_{p_1} \rangle
C^{pp'}_{h_1 p_1} +\label{eqd} \\
&& \sum_{p_1 p_2}\langle M_{p_1} M_{p_2} \rangle
(D^{pp'}_{p_1p_2} -  
{\frac {\delta_{p p_1} \delta_{p' p_2}} 
{4 (1- \langle M_{p_1} \rangle)^2 (1 - \langle M_{p_2}\rangle)^2} })  = 
F_{pp'} ~,\nonumber 
\end{eqnarray}
were we have introduced the auxiliary matrices

\ba
F_{pp'} =& - 2 \sum_{p_1}(1 - \langle M_{p_1} \rangle ) \Y_{pp_1}^2
\Y_{p'p_1}^2
+ 2 \sum_{h_1} (1- \langle M_{h_1} \rangle ) \Z_{ph_1}^2 \Z_{p'h_1}^2
\nonumber\\
&-\sum_{h_1 h_2} B^{pp'}_{h_1h_2} 
(1 -\langle M_{h_1} \rangle- \langle M_{h_2} \rangle ) \label{fpp}\\
&-\sum_{h_1 p_1} C^{pp'}_{h_1p_1} 
(1 -\langle M_{p_1} \rangle- \langle M_{h_1} \rangle ) \nonumber\\
&-\sum_{p_1 p_2} D^{pp'}_{p_1p_2} 
(1 -\langle M_{p_1} \rangle- \langle M_{p_2} \rangle ) ~,\nonumber
\ea

\ba
B^{pp'}_{h_1h_2} & = &(\Z_{ph_1}\Z_{p'h_2})^2 +
\Z_{ph_1}\Z_{p'h_1}\Z_{ph_2}\Z_{p'h_2}  ~,\nonumber \\
C^{pp'}_{h_1p_1} & = &-(\Y_{pp_1}\Z_{p'h_1} + \Y_{p'p_1}\Z_{ph_1})^2  ~,
\label{bpp}\\
D^{pp'}_{p_1p_2} & = &(\Y_{pp_1}\Y_{p'p_2})^2 +
\Y_{pp_1}\Y_{p'p_2}\Y_{p'p_1}\Y_{pp_2} ~.\nonumber 
\ea

\ba
F_{hp} =
&(\langle M_{h}\rangle + \langle M_{p} \rangle)
/(2 (1- \langle M_{h} \rangle )^2 (1- \langle M_{p} \rangle)^2 ) 
\nonumber\\
& - 2 \sum_{p_1}(1 - \langle M_{p_1} \rangle ) 
\left( \Y_{pp_1}^2 \Z_{p_1h}^2 - \X_{pp_1}^2 \Z_{hp_1}^2 \right)
\nonumber\\
&- 2 \sum_{h_1} (1 - \langle M_{h_1} \rangle )
\left(  \Z_{h_1p}^2 \Y_{hh_1}^2 - \Z_{ph_1}^2 \X_{hh_1}^2 \right)
\nonumber\\
&-\sum_{h_1 h_2} B^{ph}_{h_1h_2} 
(1 -\langle M_{h_1} \rangle- \langle M_{h_2} \rangle ) \label{fhp}\\
&-\sum_{h_1 p_1} C^{ph}_{h_1p_1} 
(1 -\langle M_{p_1} \rangle- \langle M_{h_1} \rangle ) \nonumber\\
&-\sum_{p_1 p_2} D^{ph}_{p_1p_2} 
(1 -\langle M_{p_1} \rangle- \langle M_{p_2} \rangle ) ~,\nonumber
\ea

\ba
B^{hp}_{h_1h_2} = &(\Y_{hh_1}\Z_{h_2p})^2 +
\Y_{hh_1}\Z_{h_2p}\Z_{h_1p}\Y_{hh_2} \nonumber\\
 & + (\X_{hh_1}\Z_{ph_2})^2 +
\X_{hh_1}\Z_{ph_2}\Z_{ph_1}\X_{hh_2} ~,\nonumber\\ 
C^{hp}_{h_1p_1} = &- (\Y_{pp_1}\X_{hh_1} + \Z_{p_1h}\Z_{ph_1})^2 
\label{bhp}\\
& - (\Y_{hh_1}\X_{pp_1} + \Z_{h_1p}\Z_{hp_1})^2 ~,\nonumber\\
D^{hp}_{p_1p_2} = &(\Z_{p_1h}\Y_{pp_2})^2 +
\Z_{p_1h}\Y_{pp_1}\Z_{p_2h}\Y_{pp_2}\nonumber \\
 & +(\Z_{hp_1}\X_{pp_2})^2 +
\Z_{hp_1}\X_{pp_1}\Z_{hp_2}\X_{pp_2} ~.\nonumber 
\ea

\ba
F_{hh'} =& 2 \sum_{p_1}(1 - \langle M_{p_1} \rangle ) \Z_{hp_1}^2
\Z_{h'p_1}^2
- 2 \sum_{h_1} (1-\langle M_{h_1} \rangle ) \Y_{h'h_1}^2 \Y_{hh_1}^2
\nonumber\\
&-\sum_{h_1 h_2} B^{h'h}_{h_1h_2} 
(1 -\langle M_{h_1} \rangle- \langle M_{h_2} \rangle ) \label{fhh}\\
&-\sum_{h_1 p_1} C^{h'h}_{h_1p_1} 
(1 -\langle M_{p_1} \rangle- \langle M_{h_1} \rangle ) \nonumber\\
&-\sum_{p_1 p_2} D^{h'h}_{p_1p_2} 
(1 -\langle M_{p_1} \rangle- \langle M_{p_2} \rangle )  ~,\nonumber
\ea

\ba
B^{hh'}_{h_1h_2} & = &(\Y_{h'h_1}\Y_{hh_2})^2 +
\Y_{h'h_1}\Y_{hh_1}\Y_{h'h_2}\Y_{hh_2} ~,\nonumber \\
C^{hh'}_{h_1p_1} & = &-(\Z_{h'p_1}\Y_{hh_1} + \Z_{hp_1}\Y_{h'h_1})^2 ~,
\label{bhh}\\
D^{hh'}_{p_1p_2} & = &(\Z_{h'p_1}\Z_{hp_2})^2 +
\Z_{h'p_1}\Z_{hp_2}\Z_{hp_1}\Z_{h'p_2} ~;\nonumber 
\ea

with

\ba
\X_{h h_1} = \sum\limits_{\lambda}{\frac{ X_h^{\lambda}
X_{h_1}^{\lambda}}        
{ \sqrt{(1-\langle M_{h} \rangle ) (1-\langle M_{h_1}\rangle )} }} , 
&\X_{p p_1} = \sum_{\mu}{\frac{ X_p^{\mu} X_{p_1}^{\mu}}        
{ \sqrt{(1-\langle M_{p} \rangle ) (1-\langle M_{p_1}\rangle )} }} , 
\nonumber \\
\Y_{p p_1} = \sum\limits_{\lambda}{\frac{ Y_p^{\lambda}
Y_{p_1}^{\lambda}}
{\sqrt{(1-\langle M_{p}\rangle) (1-\langle M_{p_1}\rangle)} }} ,~~~
&\Y_{h h_1} = \sum\limits_{\mu}{\frac{ Y_h^{\mu} Y_{h_1}^{\mu}}
{\sqrt{(1-\langle M_{h}\rangle) (1-\langle M_{h_1}\rangle)} }} ,
\label{XYZ} \\
Z_{p h} = \sum\limits_{\lambda}{\frac{ Y_p^{\lambda} X_h^{\lambda}}
{ \sqrt{(1-\langle M_{h}\rangle)(1-\langle M_{p}\rangle)} }},~~~
&Z_{h p} = \sum\limits_{\mu}{\frac{ Y_h^{\mu} X_p^{\mu}}
{ \sqrt{(1-\langle M_{h}\rangle)(1-\langle M_{p}\rangle)} }}.~~~
\nonumber
\ea

Due to the symmetry $\langle M_h M_p \rangle = \langle M_p M_h \rangle$
there is no need to include an explicit equation for $\langle M_p M_h
\rangle$. 

Notice that the above matrices have the following symmetry properties
\ba
B^{pp'}_{h_1h_2} = B^{p'p}_{h_2h_1}, ~~~ 
C^{pp'}_{h_1p_1} = C^{p'p}_{h_1p_1}, ~~~
D^{pp'}_{p_1p_2} = D^{p'p}_{p_2p_1},\\
B^{hh'}_{h_1h_2} = B^{h'h}_{h_2h_1}, ~~~ 
C^{hh'}_{h_1p_1} = C^{h'h}_{h_1p_1}, ~~~
D^{hh'}_{p_1p_2} = D^{h'h}_{p_2p_1}.
\ea

They imply
\be
F_{pp'} = F_{p'p}, ~~~F_{hh'} = F_{h'h}  ~,
\ee
and guaranty that $\langle M_p M_{p'} \rangle = \langle M_{p'} M_{p}
\rangle$ and $\langle M_h M_{h'} \rangle = \langle M_{h'} M_{h}
\rangle$.

Due to the particle-hole symmetry, Eqs. (\ref{x=x}) and
(\ref{m=m}), these matrices possess also the following symmetries
\be
\X_{h\,h'} = \X_{p=h\,p'=h'}, \quad
\Y_{p\,p'} = \Y_{h=p\,h'=p'}, \quad
Z_{p\,h} = Z_{h'=p\,p'=h} .
\ee
which reflects in the $A,B,C,D$ matrices as
\ba
B^{h\, h'}_{h_1\, h_2} = D^{p=h\,\, p'=h'}_{p_1=h_1\, p_2=h_2} ,\quad 
D^{h\, h'}_{p_1\, p_2} = B^{p=h\,\, p'=h'}_{h_1=p_1\, h_2=p_2} , \nonumber
\\
C^{h\, h'}_{h_1\, p_1} = C^{p=h\,\, p'=h'}_{h_1\, p_1} ,\quad
F_{p\,p'} = F_{h=p\, h'=p'},  
\ea
thus implying that the particle-hole symmetry is fulfilled in the
following sense:
\be
\langle M_{p} M_{p'} \rangle = \langle M_{h=p} M_{h'=p'} \rangle ,
\quad
\langle M_{p} M_{h} \rangle = \langle  M_{p'=h} M_{h'=p} \rangle .
\ee
These are the last two equalities which were assumed valid in (\ref{adv}).
They show that the whole formalism is consistent with the particle - hole
symmetry. With these symmetry relations Eqs. (\ref{eqb},\ref{eqc},\ref{eqd}) 
together with Eqs. (\ref{fpp},\ref{bpp},\ref{fhp},\ref{bhp},\ref{fhh},\ref{bhh}) 
can be solved numerically.

\section{Simpler RPA formalisms}

In a previous work the SCRPA equations (\ref{scrpa}) were solved under the
approximation
\be
\langle M_i  M_j \rangle  \approx 
\langle M_i \rangle \langle M_j \rangle .
\ee
This replacement close the SCRPA equations, and will be denoted SCRPA$_1$
formalism from here on. 

The renormalized RPA (r-RPA) formalism allows the consideration of the
Pauli principle in a simpler way.  The commutation relations between the
operator $Q_{i},Q_{j}^{\dagger }$ are replaced by their expectation value,
i.e. 
\be
\left[ Q_{i},Q_{j}^{\dagger }\right] \rightarrow
\delta _{ij}\left( 1-\langle M_{i}\rangle \right) .
\ee
The RPA equations take the much simpler form
\ba
A^{\hbox{\footnotesize r-RPA}}_{pp'} =& 2 \delta_{pp'} \left(\epsilon (p - {\frac 1 2}) +
{\frac G 2} \right)
- G \sqrt{(1 - \langle M_p \rangle) (1 -\langle  M_{p'}\rangle)} 
\nonumber \\
B^{\hbox{\footnotesize r-RPA}}_{ph} =&  G \sqrt{(1 - \langle M_p \rangle) (1 -\langle  M_{h}\rangle)}
\label{rrpa}\\
C^{\hbox{\footnotesize r-RPA}}_{hh'} =& -2 \delta_{hh'} \left(\epsilon (h - {\frac 1 2})
+  {\frac G 2} \right)
+ G \sqrt{(1 - \langle M_h \rangle) (1 -\langle  M_{h'}\rangle)}
\nonumber 
\ea
They can be seen as a simplification of the SCRPA$_1$ equations when the
limit
\be
\langle Q^\dagger_i Q_j \rangle \rightarrow 0 , \quad \quad
\langle Q_i Q_j \rangle \rightarrow 0 .
\ee
is assumed valid. 
At the same time, we see that the correlation energy can only be obtained
using the expression for $E^{RPA}_{corr}$.

Taking additionally the limit
\be
\langle M_i \rangle \rightarrow 0 ,
\ee
the standard RPA matrices for the picket fence model
\ba
A^{RPA}_{pp'} =& 2 \delta_{pp'} \left(\epsilon (p - {\frac 1 2}) +
{\frac G 2} \right) - G ~,
\nonumber \\
B^{RPA}_{ph} =&  G  ~,\label{rpa-mat}\\
C^{RPA}_{hh'} =& -2 \delta_{hh'} \left(\epsilon (h - {\frac 1 2})
+  {\frac G 2} \right) + G ~,
\nonumber 
\ea
are recovered. 

\section{Sum rules}

Given that the present RPA formalism is number conserving, we know that

\be
N|RPA \rangle = \Omega |RPA \rangle, \hbox{~~with~~} 
N = \sum_h N_h + \sum_p N_p ~,
\ee
which satisfy
\be
\langle N \rangle = \sum\limits_{h=1}^{\Omega/2} (2 - \langle M_h \rangle)
+ \sum\limits_{p=1}^{\Omega/2} 
\langle M_p \rangle = \Omega + \sum\limits_{i=1}^{\Omega/2} (\langle 
M_{p=i} \rangle - \langle M_{h=i} \rangle ) 
= \Omega 
\ee
due to (\ref{m=m}).

Given that
\ba
 \Omega^2 = \langle N^2 \rangle  
= \langle (\Omega + \sum\limits_i (M_{p=i} - M_{h=i}))
(\Omega + \sum\limits_j (M_{p=j} - M_{h=j})) \rangle \nonumber\\
= \Omega^2 + 2 \Omega  \sum\limits_i (M_{p=i} - M_{h=i})  \\
+ \sum\limits_{ij} \left( \langle M_{p=i} M_{p=j} \rangle
+ \langle M_{h=i} M_{h=j} \rangle - \langle M_{p=i} M_{h=j} \rangle
- \langle M_{h=i} M_{p=j} \rangle \right)\nonumber 
\ea
we get an additional condition over the expectation values, which is

\be
\sum\limits_{pp'} \langle M_{p} M_{p'} \rangle 
= \sum\limits_{hp} \langle M_{h} M_{p} \rangle . \label{sum}
\ee
This sum rule plays an important role in the discussions of the following
section.

\section{The SCRPA ground state}

In all our previous derivations we always made use of the vacuum condition
(\ref{vac-rpa}) for the ground state $| SCRPA \rangle$ of the present theory.
However, at no step we have constructed the ground state explicitly. This
situation is in fact quit common. Also Hartree-Fock and 
Hartree-Fock-Bogoliubov theories can be derived using analogous vacuum conditions,
without constructing the ground state wave function explicitly (in spite
of the fact that their construction is perfectly possible).
The question then naturally arises whether from (\ref{vac-rpa}) it is
possible to explicitly construct the SCRPA ground state.

To this purpose let us first consider the $\Omega = 2$ case.
In this case there is only one particle state ($p$) and one hole state ($h$).
There are two unperturbed (basis) states with 2 particles. They are
\be
|h\rangle = |HF\rangle \quad \hbox{and} \quad
|p\rangle = Q^\dagger_p Q^\dagger_h |HF\rangle  .
\ee
The first state has the two particles in the lower state ($h$), the
second in the upper state ($p$). 

The SCRPA ground state can be expanded in this basis as
\be
| SCRPA \rangle = a_h |h\rangle + a_p |p\rangle \quad \hbox{with} \quad
a_h^2 + a_p^2 = 1 .
\ee
The coefficients $a_h, a_p$ must be determined from the vacuum condition
(\ref{vac-rpa}), which in this simple case reads
\be
A \,| SCRPA \rangle = ( X_p \overline{Q}^\dagger_p - Y_h \overline{Q}_h )
\, | SCRPA \rangle = 0  . 
\ee
In this case $X_p = X_h = X$ and $Y_p = Y_h = Y$, and the removal operator
$R$ acting on the SCRPA vacuum yields the same equation, which is
\be
X a_p - Y a_h = 0 ~~.
\ee
Together with the normalization condition, they have the solutions
\be
a_h = {\frac {X} {\sqrt{X^2 + Y^2}} }~~,~~~~~
a_p = {\frac {Y} {\sqrt{X^2 + Y^2}} }~~.
\ee
In this way the SCRPA ground state is entirely determined by the 
RPA amplitudes X,Y. In the following section it will in fact be shown that
the SCRPA can reproduces exactly the correlation and excitation energies
in the case of $\Omega = 2$.

Let us now consider the $\Omega = 4$ case. 
The phonon operators for the addition mode are
\be
A_{\mu }^{\dagger }=\sum_{p=1}^2 X_{p}^{\mu }\ \overline{Q}_{p}^{\dagger
}-\sum_{h=1}^2 Y_{h}^{\mu }\ \overline{Q}_{h}
=\sum_{p=1}^2 \x_{p}^{\mu }  Q_{p}^{\dagger
}-\sum_{h=1}^2 \y_{h}^{\mu }  Q_{h} ,\label{amu}
\ee
were we have introduced the shorthand notation
\be
\x_{p}^{\mu } = X_{p}^{\mu } /\sqrt{1-\langle M_p \rangle},\hbox{~~~~~~~}
\y_{h}^{\mu } = Y_{h}^{\mu } /\sqrt{1 -\langle M_h \rangle}~.
\ee

The SCRPA vacuum for 4 particles states has the form
\be
| SCRPA \rangle = a_0 \,|HF\rangle + \sum_{hp} a_{hp}\, 
Q^\dagger_{h} Q^\dagger_{p} | HF\rangle +
a_2 \,Q^\dagger_{h_1} Q^\dagger_{h_2} Q^\dagger_{p_1}
Q^\dagger_{p_2} | HF\rangle .\label{vac}
\ee
The six coefficients $a_0, a_{hp}, a_2$ must be
determined using the normalization and the vacuum conditions. 
They yield to the equations
\ba
\sum_p \x^\mu_p a_{hp} - \y^\mu_{h} a_{0} = 0 ~~,
\qquad
\x^\mu_{p} a_{2} - \sum_h \y^\mu_{h} a_{hp} = 0  ~~.
\label{eqpp}
\ea

There are four of these equations (2 for $h = 1,2$ and two for
$p = 1, 2$) for each of the 2 possible $\mu$'s, giving a total of 8
equations, plus the normalization condition, to obtain the 6 $a$'s.

These equations are consistent with the simplest HF (G=0) limit.
In this case
\be
\x^\mu_p = \delta_{\mu p}, \hbox{~~~~} \y^\mu_{h} = 0 ~,\hbox{~~~~} 
\mu, p, h =1,2
\ee
From Eq. (\ref{eqpp}) we get $a_{hp} = 0$ for all $h,p$, and 
$a_{2} =0$. The only remaining variable to be
determined is $a_{0} = 1.0$ from the normalization condition.
The HF ground state is recovered as expected.

Given that the $\x^\mu_{p}$'s are always finite, we can obtain four
different expressions for $a_{2}$. They are
\ba
a_{2} = {\frac {\y^1_{1}}{\x^1_{2}} } a_{1 1}
	    + {\frac {\y^1_{2}}{\x^1_{2}} } a_{2 1}  
	    = {\frac {\y^1_{1}}{\x^1_{1}} } a_{1 2}
	    + {\frac {\y^1_{2}}{\x^1_{1}} } a_{2 2}
\nonumber \\
 	    = {\frac {\y^2_{1}}{\x^2_{2}} } a_{1 1}
	    + {\frac {\y^2_{2}}{\x^2_{2}} } a_{2 1}
 	    = {\frac {\y^2_{1}}{\x^2_{1}} } a_{1 2}
	    + {\frac {\y^2_{2}}{\x^2_{1}} } a_{2 2}
\ea

They imply 
\ba
{\frac {\y^1_{1}}{\x^1_{1}} } =
{\frac {\y^2_{1}}{\x^2_{1}} } ~,
\hbox{~~~~~~~~~}
{\frac {\y^1_{2}}{\x^1_{1}} } =
{\frac {\y^2_{2}}{\x^2_{1}} } ~,
\nonumber \\
{\frac {\y^1_{1}}{\x^1_{2}} } =
{\frac {\y^2_{1}}{\x^2_{2}} } ~,
\hbox{~~~~~~~~~}
{\frac {\y^1_{2}}{\x^1_{2}} } =
{\frac {\y^2_{2}}{\x^2_{2}} } ~~.\label{xy}
\ea
In general there is no reason for relations (\ref{xy}) to be
fulfilled. Indeed, in numerical examples they are not satisfied
for any finite value of $G$. For larger spaces the situation
does not improve. It implies that besides for $\Omega = 2$, the
SCRPA ground state defined by Eq. (\ref{vac-rpa}) does not
in fact exist. However, the theory is perfectly consistent in itself and
(\ref{vac-rpa}) must therefore be interpreted as an auxiliary relation
which allows to express expectation values such as in (\ref{qq}) and 
(\ref{m}) in a well defined way in terms of the RPA amplitudes.

In fact, using for example the Green's function method, as described in
\cite{Duk98a}, one can obtain expressions (\ref{qq},\ref{m}) via the residua
of the Green's functions without ever using the vacuum condition 
(\ref{vac-rpa}) explicitly. On the other hand, the non-existence of $| SCRPA
\rangle$ entails that we are not assured that our theory yields an upper 
bound of the ground state energy as in a truly Raleigh-Ritz variational
principle. It also implies that our theory eventually contains some violation
of the Pauli principle, in spite of the fact that we never made use of any
boson approximation and always fully respected the commutation laws of the
Fermi pair operators. In the next section, where we present the numerical
results, we will return and further discuss some consequences of the 
non-existence of the SCRPA vacuum.

\section{Numerical solutions}

In the present section the picket fence model is solved using the 
fully Self-Consistent RPA for different number of levels  $\Omega$.
The ground state and excited states energies, as well as the particle
numbers and correlation functions, are compared with
the exact ones, and with those obtained from the SCRPA$_1$, the
r-RPA and the RPA.
For $\Omega = 2$ analytical results are presented.
In all the cases we fix the energy scale by setting
$\epsilon = 1 .$

\subsection{The $\Omega=2$ case}

This case was introduced in the previous section..
The Fermi level is at the energy
\be
\lambda = {\frac 3 2} - {\frac G 2}
\ee

Exact correlated states are built as linear combinations of the states
basis states $|h\rangle$ and $|p\rangle$:
\be
|\alpha \rangle = c_h |h\rangle + c_p |p\rangle \quad \hbox{with} \quad
c_p^2 + c_h^2 = 1 ,
\ee
which satisfy the eigenvalue equation

\be
\left( \begin{array}{cc} -1 &G \\ G &1 \end{array} \right )
\left( \begin{array}{c} c_h \\ c_p \end{array} \right )
= E_\alpha 
\left( \begin{array}{c} c_h \\ c_p \end{array} \right )
\ee
with solutions
\be
E_\pm = \pm \sqrt{1+ G^2} , \quad\quad
c_h = {\frac {G} {\sqrt{G^2 +(E_\pm +1)^2}} } , \quad
c_p = {\frac {E_\pm +1} {\sqrt{G^2 +(E_\pm +1)^2}} }.
\ee

The exact ground state $|0\rangle$ has energy $E_- = -\sqrt{1+ G^2}$.
The correlation energy $E_{\hbox{corr}}$ is
\be
E_{\hbox{corr}} = \langle 0| H |0\rangle - \langle HF| H |HF\rangle =
-\sqrt{1+ G^2} + 1  ~.
\ee

The exact particle number expectation values are
\be
\langle 0| M_h | 0\rangle = 
\langle 0| M_p | 0\rangle = 2 a_2 ,
\ee
and the exact correlation functions are
\be
\langle 0| M_h M_h | 0\rangle = 
\langle 0| M_p M_p | 0\rangle =
\langle 0| M_p M_h | 0\rangle = 4 a_1^2
\ee

The addition mode has 4 particles and only one state
\be
|4\rangle = Q^\dagger_p |HF \rangle ~,
\ee
with
\be
\langle 4 |H |4 \rangle = 0 ~. 
\ee

The exact excitation energy associated to this mode is
\be
E_{\hbox{exc}} = \langle 4 |H |4\rangle - \langle 0 |H |0 \rangle  =
-E_- = \sqrt{1+ G^2} .
\ee

The SCRPA equations for the addition mode are

\be
\left( \begin{array}{cc} 1- 2 GXY &G(1+2Y^2) \\-G(1+2Y^2)  &-1+ 2 GXY 
\end{array} \right )
\left( \begin{array}{c} X \\ Y \end{array} \right )
= E
\left( \begin{array}{c} X \\ Y \end{array} \right )
\ee
with solutions
\be
E_{1,2} = \pm \sqrt{1+ G^2} , \quad\quad
X = {\frac {G} {\sqrt{G^2 -(1- E_{1,2})^2}} }, \quad
Y = {\frac {1-E_{1,2}} {\sqrt{G^2 - (1- E_{1,2})^2}} }.
\ee
The addition mode has excitation energy $E_1$, which is exactly the same
as $E_{\hbox{exc}}$ deduced above, although the expressions for $X, Y$ are
quite different from those obtained for $a_1, a_2$. 

The occupation numbers are

\be
\langle M_p \rangle = \langle M_h \rangle = 1 - {\frac {1} {1+2Y^2}} =
2 {\frac {(1-E_{1})^2} {G^2 - (1- E_{1})^2}} 
\ee
which again reproduce the exact results.

The  standard RPA equations for the addition mode are

\be
\left( \begin{array}{cc} 1 &G \\-G  &-1 
\end{array} \right )
\left( \begin{array}{c} X \\ Y \end{array} \right )
= E_{RPA}
\left( \begin{array}{c} X \\ Y \end{array} \right )
\ee
with solutions
\be
E_{RPA} = \pm \sqrt{1- G^2} 
\ee
The RPA excitation energy exhibits the well known ``collapse'' when
$G = 1$, and is a decreasing function of G, while the exact energy is an
increasing function.

The  renormalized RPA (r-RPA) equations for the addition mode are

\be
\left( \begin{array}{cc}
 1 + G {\frac {2 Y^2}{1+2Y^2}} & G {\frac {1} {1+2Y^2}}   
\\-G {\frac {1} {1+2Y^2}}  &-1 - G {\frac {2 Y^2}{1+2Y^2}} 
\end{array} \right )
\left( \begin{array}{c} X \\ Y \end{array} \right )
= E_{r-RPA}
\left( \begin{array}{c} X \\ Y \end{array} \right )  .
\ee
Its solutions are not analytically simple. Numerical results can be seen in
Tables \ref{2.0} and \ref{2.1}, together with the energies obtained
with the other formalisms and the exact solutions. 

The contrast with the standard RPA exhibits the power of the fully self
consistent RPA formalism, which in the $\Omega = 2$ case was shown to
reproduce the exact results.

\begin{table}
\begin{tabular}{llllll}
G  & exact & RPA  & r-RPA & SCRPA$_1$  & SCRPA \\
\hline
0.0  &  \,0.0  &    \,0.0   &   \,0.0  &  \,0.0    & \,0.0    \\
0.2  & -0.0198 & -0.0202 & -0.0198 & -0.0198 & -0.0198 \\
0.4  & -0.0770 & -0.0835 & -0.0770 & -0.0755 & -0.0770 \\
0.6  & -0.1661 & -0.2000 & -0.1362 & -0.1568 & -0.1661 \\
0.8  & -0.2806 & -0.4000 & -0.2065 & -0.2532 & -0.2806 \\
0.90 & -0.3454 & -0.5641 & -0.2415 & -0.3050 & -0.3454  \\
\end{tabular}

\caption{Correlation energies as function of G for
$\Omega = 2$ obtained with an exact
calculation, with the RPA, r-RPA, SCRPA$_1$ and SCRPA methods.} 
\label{2.0}
\end{table}
\begin{table}
\begin{tabular}{llllll}
G  & exact & RPA  & r-RPA & SCRPA$_1$  & SCRPA \\
\hline
0.0  & 1.0  &  1.0   &  1.0  & 1.0    & 1.0    \\
0.2  & 1.0198  & 0.9798 & 0.9845 & 1.0045 & 1.0198 \\
0.4  & 1.0770  & 0.9165 & 0.9572 & 1.0139 & 1.0770 \\
0.6  & 1.1662  & 0.8000 & 0.9397 & 1.0227 & 1.1662 \\
0.8  & 1.2806  & 0.6000 & 0.9379 & 1.4155 & 1.2806 \\
0.90 & 1.3454  & 0.4359 & 0.9419 & 1.5013 & 1.3454 \\
\end{tabular}

\caption{Energies of the first excited state as function of G for
$\Omega = 2$ obtained
with an exact calculation, with the RPA, r-RPA,  SCRPA$_1$ and SCRPA methods.} 
\label{2.1}
\end{table}

\subsection{The $\Omega=4$ case}

We label the four levels, from bottom to top, as $h_2, h_1, p_1,
p_2$.  
The exact eigenstates for 4 particles states, including the ground
state, are 
\be
|\alpha\rangle = c^\alpha_0 |HF\rangle + \sum_{hp} c^\alpha_{hp} 
Q^\dagger_{h} Q^\dagger_{p} | HF\rangle +
c^\alpha_2 Q^\dagger_{h_1} Q^\dagger_{h_2} Q^\dagger_{p_1}
Q^\dagger_{p_2} | HF\rangle .
\ee
The coefficients $c^\alpha_{\mu}$ fulfill the normalization condition
\be
\sum_{\mu} |c^\alpha_{\mu}|^2 = 1 ~,
\ee
and are obtained as the eigen vectors of 

\be
\left( \begin{array}{cccccc}
-4 & -G & -G & -G & -G & 0  \\ 
-G & -2 & -G & -G &  0 & -G \\ 
-G & -G &  0 &  0 & -G & -G \\ 
-G & -G &  0 &  0 & -G & -G \\ 
-G &  0 & -G & -G &  2 & -G \\ 
 0 & -G & -G & -G & -G & 4 \\ 
\end{array} \right )
\left( \begin{array}{c} 
c^\alpha_0 \\ c^\alpha_{h_1 p_1} \\ c^\alpha_{h_1 p_2}\\ c^\alpha_{h_2
p_1} \\ c^\alpha_{h_2 p_2} \\ c^\alpha_2
\end{array} \right )
= E_\alpha
\left( \begin{array}{c} 
c^\alpha_0 \\ c^\alpha_{h_1 p_1} \\ c^\alpha_{h_1 p_2}\\ c^\alpha_{h_2
p_1} \\ c^\alpha_{h_2 p_2} \\ c^\alpha_2
\end{array} \right ) ~.
\ee

The particle-hole symmetry implies that there are two degenerate
unperturbed states, which are
\be
Q^\dagger_{h_1} Q^\dagger_{p_2} | HF\rangle \quad \hbox{and} \quad
Q^\dagger_{h_2} Q^\dagger_{p_1} | HF\rangle , 
\ee
both having zero unperturbed energies. They always contribute on equal
footing to any correlated state, i.e.
\be
c^\alpha_{h_1 p_2} = c^\alpha_{h_2 p_1} . \label{c=c}
\ee

The exact occupation numbers are
\ba
\langle \alpha | M_h | \alpha \rangle =
2 \left( |c^\alpha_{h p_1}|^2 + |c^\alpha_{h p_2}|^2 + |c^\alpha_2|^2
\right) , \nonumber \\
\langle \alpha | M_p | \alpha \rangle =
2 \left( |c^\alpha_{h_1 p}|^2 + |c^\alpha_{h_2 p}|^2 + |c^\alpha_2|^2
\right) .
\ea
which due to (\ref{c=c}) imply $\langle \alpha | M_h | \alpha \rangle 
= \langle \alpha | M_{p=h} | \alpha \rangle $;
and
\ba
\langle \alpha | M_h M_h | \alpha \rangle =
2 \, \langle \alpha | M_h | \alpha \rangle , &
\langle \alpha | M_p M_p | \alpha \rangle =
2 \, \langle \alpha | M_p | \alpha \rangle , \nonumber \\
\langle \alpha | M_{h_1} M_{h_2} | \alpha \rangle = 4 \, |c^\alpha_2|^2
, & 
\langle \alpha | M_{p_1} M_{p_2} | \alpha \rangle = 4 \,
|c^\alpha_2|^2 ~,\\
\langle \alpha | M_{p} M_{h} | \alpha \rangle = 
4 \, |c^\alpha_{h p}|^2  + 4 \, |c^\alpha_2|^2 ~.\nonumber
\ea

The exact eigenstates for states with 6 particles (addition mode) are 
\be
|\beta\rangle = \sum_{p} c^\beta_{p} Q^\dagger_{p} | HF\rangle +
\sum_{h} c^\beta_{h}  Q^\dagger_{h} Q^\dagger_{p_1} Q^\dagger_{p_2} |
HF\rangle .\label{exc}
\ee
The coefficients $c^\beta_{p,h}$ are obtained as the eigen vectors of 

\be
\left( \begin{array}{cccc}
-3 & -G & G & G  \\ 
-G & -1 & G & G  \\ 
 G &  G &  1 & -G \\ 
 G &  G &  -G &  3 
\end{array} \right )
\left( \begin{array}{c} 
c^\beta_{p_1} \\ c^\beta_{p_2}\\ c^\beta_{h_1} \\ c^\beta_{h_2} 
\end{array} \right )
= E_\beta
\left( \begin{array}{c} 
c^\beta_{p_1} \\ c^\beta_{p_2}\\ c^\beta_{h_1} \\ c^\beta_{h_2} 
\end{array} \right )
\ee

The exact excitation $E^\beta_{\hbox{exc}}$ energies for the addition mode
are measured against the ground state energy, i.e.
\be
E^\beta_{\hbox{exc}} = E_\beta - E_{\alpha =1} . 
\ee

We have calculated the exact solutions, and compared them with the
solutions of the SCRPA equations. This involved to solve simultaneously
the SCRPA eigenvalue problem, Eqs.(\ref{rpa}) and to evaluate the number 
correlations by solving Eq. (\ref{mm}). In the following tables we present
the correlation energies and the excitation energies of the first
and second addition modes, obtained using four different methods: the
exact results, the RPA ones, the SCRPA using $\langle M_i M_j \rangle =
\langle M_i \rangle \langle M_j \rangle$ (denoted SCRPA$_1$) and the fully
Self-Consistent RPA (SCRPA) which refers to the
treatment described above.

\begin{table}
\begin{tabular}{llllll}
G  & exact & RPA  & r-RPA  & SCRPA$_1$  & SCRPA \\
\hline
0.0  &  \,0.0  &\,0.0    &  \,0.0  & \,0.0   & \,0.0  \\
0.1  & -0.0123 & -0.0124 & -0.0123 & -0.0124 & -0.0124 \\
0.2  & -0.0520 & -0.0538 & -0.0510 & -0.0523 & -0.0524 \\
0.3  & -0.1225 & -0.1343 & -0.1158 & -0.1233 & -0.1248 \\
0.4  & -0.2264 & -0.2764 & -0.2009 & -0.2259 & -0.2334 \\
0.5  & -0.3644 & -0.5662 & -0.2961 & -0.3575 & -0.3795
\end{tabular}

\caption{Correlation energies as function of G for
$\Omega = 4$ obtained with exact
calculation, with the RPA, r-RPA, SCRPA$_1$ and SCRPA methods.} 
\label{4.0}
\end{table}

\begin{table}
\begin{tabular}{llllll}
G  & exact & RPA  & r-RPA  & SCRPA$_1$  & SCRPA \\
\hline
0.0  & 1.0    &   1.0  &   1.0  &  1.0    & 1.0   \\
0.1  & 1.0027  & 0.9864 & 0.9872 & 1.0041 & 1.0027  \\
0.2  & 1.0115  & 0.9402 & 0.9494 & 1.0233 & 1.0126 \\
0.3  & 1.0276  & 0.8493 & 0.8917 & 1.0648 & 1.0348 \\
0.4  & 1.0520  & 0.6898 & 0.8248 & 1.1279 & 1.0734 \\
0.5  & 1.0853  & 0.3745 & 0.7603 & 1.2065 & 1.1274
\end{tabular}

\caption{Excitation energy of the first addition mode as function of G
for $\Omega = 4$ obtained with exact
calculation, with the RPA, r-RPA, SCRPA$_1$ and SCRPA methods.} 
\label{4.1}
\end{table}

\begin{table}
\begin{tabular}{llllll}
G  & exact & RPA  & r-RPA  & SCRPA$_1$  & SCRPA \\
\hline
0.0  & 3.0    &   3.0  &  3.0  &  3.0   & 3.0    \\
0.1  & 3.0104  & 3.0012 & 3.0014 & 3.0108 & 3.0104 \\ 
0.2  & 3.0465  & 3.0060 & 3.0075 & 3.0499 & 3.0468 \\
0.3  & 3.1156  & 3.0164 & 3.0212 & 3.1259 & 3.1168 \\
0.4  & 3.2237  & 3.0338 & 3.0446 & 3.2422 & 3.2256 \\
0.5  & 3.3738  & 3.0594 & 3.0784 & 3.3967 & 3.3784
\end{tabular}

\caption{Excitation energy of the second addition mode as function of G
for $\Omega = 4$ obtained with exact
calculation, with the RPA, r-RPA, SCRPA$_1$ and SCRPA methods.} 
\label{4.2}
\end{table}

In the three tables \ref{4.0},\ref{4.1},\ref{4.2} the improvement
offered by the SCRPA as compared
with the RPA is clearly seen. Both for the correlation and the excitation
energies the RPA gives only a crude approximation, while the SCRPA results
are very close to the exact ones.

The correlation energies listed in Table \ref{4.0} are slightly better
described
with the SCRPA$_1$ than with the SCRPA. It is also noticeably that both
the RPA and SCRPA correlation energies are lower than the exact ones,
but the SCRPA$_1$ energies are slightly higher.
The situation is very different when excitation energies are analyzed
(Tables \ref{4.1} and \ref{4.2}).
While  again the RPA energies are lower than the exact ones, both the
SCRPA and SCRPA$_1$ energies are higher. As mentioned above, they are very
close to the exact ones, but in this case the SCRPA ones are clearly the
best, reproducing with high accuracy the exact results. 

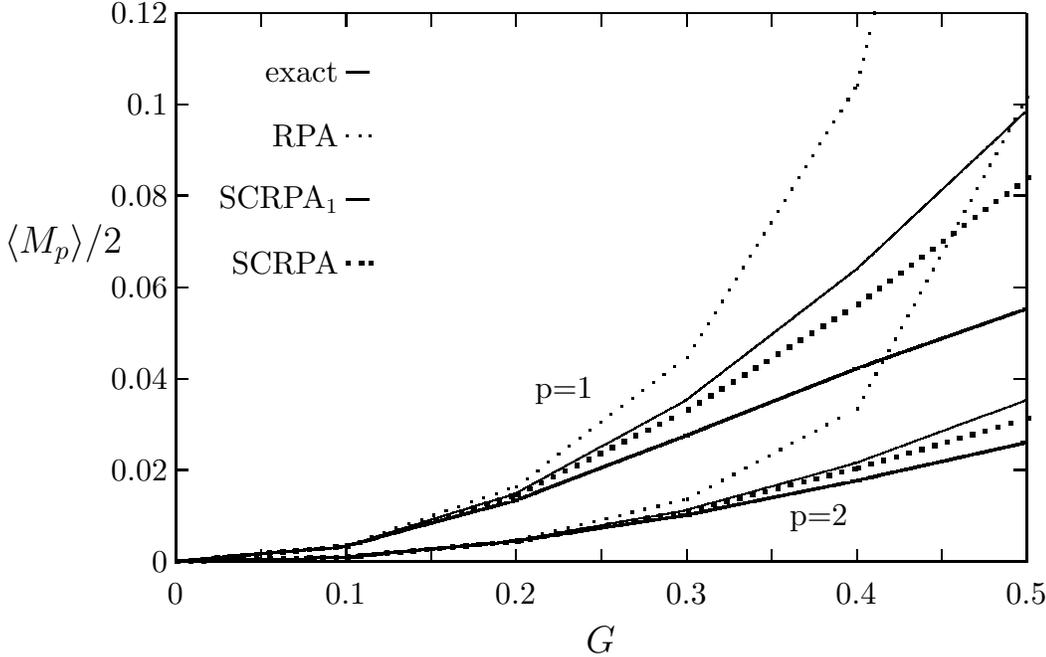
\begin{figure}
\setlength{\unitlength}{0.240900pt}
\ifx\plotpoint\undefined\newsavebox{\plotpoint}\fi
\begin{picture}(1500,900)(0,0)
\font\gnuplot=cmr10 at 12pt
\gnuplot
\sbox{\plotpoint}{\rule[-0.400pt]{0.800pt}{0.800pt}}%
\put(120.0,31.0){\rule[-0.200pt]{321.842pt}{0.400pt}}
\put(120.0,31.0){\rule[-0.200pt]{0.400pt}{207.656pt}}
\put(120.0,31.0){\rule[-0.200pt]{4.818pt}{0.400pt}}
\put(108,31){\makebox(0,0)[r]{0}}
\put(1436.0,31.0){\rule[-0.200pt]{4.818pt}{0.400pt}}
\put(120.0,175.0){\rule[-0.200pt]{4.818pt}{0.400pt}}
\put(108,175){\makebox(0,0)[r]{0.02}}
\put(1436.0,175.0){\rule[-0.200pt]{4.818pt}{0.400pt}}
\put(120.0,318.0){\rule[-0.200pt]{4.818pt}{0.400pt}}
\put(108,318){\makebox(0,0)[r]{0.04}}
\put(1436.0,318.0){\rule[-0.200pt]{4.818pt}{0.400pt}}
\put(120.0,462.0){\rule[-0.200pt]{4.818pt}{0.400pt}}
\put(108,462){\makebox(0,0)[r]{0.06}}
\put(1436.0,462.0){\rule[-0.200pt]{4.818pt}{0.400pt}}
\put(120.0,606.0){\rule[-0.200pt]{4.818pt}{0.400pt}}
\put(108,606){\makebox(0,0)[r]{0.08}}
\put(1436.0,606.0){\rule[-0.200pt]{4.818pt}{0.400pt}}
\put(120.0,749.0){\rule[-0.200pt]{4.818pt}{0.400pt}}
\put(108,749){\makebox(0,0)[r]{0.1}}
\put(1436.0,749.0){\rule[-0.200pt]{4.818pt}{0.400pt}}
\put(120.0,893.0){\rule[-0.200pt]{4.818pt}{0.400pt}}
\put(108,893){\makebox(0,0)[r]{0.12}}
\put(1436.0,893.0){\rule[-0.200pt]{4.818pt}{0.400pt}}
\put(120.0,31.0){\rule[-0.200pt]{0.400pt}{4.818pt}}
\put(120,-19){\makebox(0,0){0}}
\put(120.0,873.0){\rule[-0.200pt]{0.400pt}{4.818pt}}
\put(254.0,31.0){\rule[-0.200pt]{0.400pt}{4.818pt}}
\put(254.0,873.0){\rule[-0.200pt]{0.400pt}{4.818pt}}
\put(387.0,31.0){\rule[-0.200pt]{0.400pt}{4.818pt}}
\put(387,-19){\makebox(0,0){0.1}}
\put(387.0,873.0){\rule[-0.200pt]{0.400pt}{4.818pt}}
\put(521.0,31.0){\rule[-0.200pt]{0.400pt}{4.818pt}}
\put(521.0,873.0){\rule[-0.200pt]{0.400pt}{4.818pt}}
\put(654.0,31.0){\rule[-0.200pt]{0.400pt}{4.818pt}}
\put(654,-19){\makebox(0,0){0.2}}
\put(654.0,873.0){\rule[-0.200pt]{0.400pt}{4.818pt}}
\put(788.0,31.0){\rule[-0.200pt]{0.400pt}{4.818pt}}
\put(788.0,873.0){\rule[-0.200pt]{0.400pt}{4.818pt}}
\put(922.0,31.0){\rule[-0.200pt]{0.400pt}{4.818pt}}
\put(922,-19){\makebox(0,0){0.3}}
\put(922.0,873.0){\rule[-0.200pt]{0.400pt}{4.818pt}}
\put(1055.0,31.0){\rule[-0.200pt]{0.400pt}{4.818pt}}
\put(1055.0,873.0){\rule[-0.200pt]{0.400pt}{4.818pt}}
\put(1189.0,31.0){\rule[-0.200pt]{0.400pt}{4.818pt}}
\put(1189,-19){\makebox(0,0){0.4}}
\put(1189.0,873.0){\rule[-0.200pt]{0.400pt}{4.818pt}}
\put(1322.0,31.0){\rule[-0.200pt]{0.400pt}{4.818pt}}
\put(1322.0,873.0){\rule[-0.200pt]{0.400pt}{4.818pt}}
\put(1456.0,31.0){\rule[-0.200pt]{0.400pt}{4.818pt}}
\put(1456,-19){\makebox(0,0){0.5}}
\put(1456.0,873.0){\rule[-0.200pt]{0.400pt}{4.818pt}}
\put(120.0,31.0){\rule[-0.200pt]{321.842pt}{0.400pt}}
\put(1456.0,31.0){\rule[-0.200pt]{0.400pt}{207.656pt}}
\put(120.0,893.0){\rule[-0.200pt]{321.842pt}{0.400pt}}
\put(-56,530){\makebox(0,0){\large  $\langle M_p \rangle /{2} $}}
\put(788,-91){\makebox(0,0){\large $G$}}
\put(120.0,31.0){\rule[-0.200pt]{0.400pt}{207.656pt}}
\put(376,800){\makebox(0,0)[r]{exact}}
\put(388,800){\rule[-0.200pt]{8.672pt}{0.400pt}}
\put(120,31){\usebox{\plotpoint}}
\multiput(120.00,31.58)(5.639,0.496){45}{\rule{4.550pt}{0.120pt}}
\multiput(120.00,30.17)(257.556,24.000){2}{\rule{2.275pt}{0.400pt}}
\multiput(387.00,55.58)(1.613,0.499){163}{\rule{1.387pt}{0.120pt}}
\multiput(387.00,54.17)(264.122,83.000){2}{\rule{0.693pt}{0.400pt}}
\multiput(654.00,138.58)(0.912,0.499){291}{\rule{0.829pt}{0.120pt}}
\multiput(654.00,137.17)(266.279,147.000){2}{\rule{0.415pt}{0.400pt}}
\multiput(922.00,285.58)(0.651,0.500){407}{\rule{0.621pt}{0.120pt}}
\multiput(922.00,284.17)(265.711,205.000){2}{\rule{0.310pt}{0.400pt}}
\multiput(1189.00,490.58)(0.536,0.500){495}{\rule{0.529pt}{0.120pt}}
\multiput(1189.00,489.17)(265.902,249.000){2}{\rule{0.264pt}{0.400pt}}
\put(376,700){\makebox(0,0)[r]{RPA}}
\multiput(388,700)(20.756,0.000){2}{\usebox{\plotpoint}}
\put(424,700){\usebox{\plotpoint}}
\put(120,31){\usebox{\plotpoint}}
\multiput(120,31)(20.665,1.935){13}{\usebox{\plotpoint}}
\multiput(387,56)(19.601,6.827){14}{\usebox{\plotpoint}}
\multiput(654,149)(16.545,12.532){16}{\usebox{\plotpoint}}
\multiput(922,352)(10.986,17.610){25}{\usebox{\plotpoint}}
\multiput(1189,780)(4.824,20.187){5}{\usebox{\plotpoint}}
\put(1216,893){\usebox{\plotpoint}}
\put(376,600){\makebox(0,0)[r]{SCRPA$_1$}}
\put(388.0,600){\rule[-0.400pt]{8.672pt}{0.800pt}}
\put(120,31){\usebox{\plotpoint}}
\multiput(120.00,32.41)(5.731,0.504){41}{\rule{9.100pt}{0.122pt}}
\multiput(120.00,29.34)(248.112,24.000){2}{\rule{4.550pt}{0.800pt}}
\multiput(387.00,56.41)(1.867,0.501){137}{\rule{3.167pt}{0.121pt}}
\multiput(387.00,53.34)(260.427,72.000){2}{\rule{1.583pt}{0.800pt}}
\multiput(654.00,128.41)(1.319,0.501){197}{\rule{2.302pt}{0.121pt}}
\multiput(654.00,125.34)(263.222,102.000){2}{\rule{1.151pt}{0.800pt}}
\multiput(922.00,230.41)(1.276,0.501){203}{\rule{2.234pt}{0.121pt}}
\multiput(922.00,227.34)(262.363,105.000){2}{\rule{1.117pt}{0.800pt}}
\multiput(1189.00,335.41)(1.426,0.501){181}{\rule{2.472pt}{0.121pt}}
\multiput(1189.00,332.34)(261.869,94.000){2}{\rule{1.236pt}{0.800pt}}
\sbox{\plotpoint}{\rule[-0.800pt]{1.600pt}{1.600pt}}%
\put(376,500){\makebox(0,0)[r]{SCRPA}}
\multiput(388,500)(20.756,0.000){2}{\usebox{\plotpoint}}
\put(424,500){\usebox{\plotpoint}}
\put(120,31){\usebox{\plotpoint}}
\multiput(120,31)(20.672,1.858){13}{\usebox{\plotpoint}}
\multiput(387,55)(19.862,6.025){14}{\usebox{\plotpoint}}
\multiput(654,136)(18.592,9.227){14}{\usebox{\plotpoint}}
\multiput(922,269)(17.627,10.959){15}{\usebox{\plotpoint}}
\multiput(1189,435)(16.642,12.403){16}{\usebox{\plotpoint}}
\put(1456,634){\usebox{\plotpoint}}
\sbox{\plotpoint}{\rule[-0.400pt]{0.800pt}{0.800pt}}%
\put(120,31){\usebox{\plotpoint}}
\multiput(120.00,31.59)(20.325,0.485){11}{\rule{15.357pt}{0.117pt}}
\multiput(120.00,30.17)(235.125,7.000){2}{\rule{7.679pt}{0.400pt}}
\multiput(387.00,38.58)(5.199,0.497){49}{\rule{4.208pt}{0.120pt}}
\multiput(387.00,37.17)(258.267,26.000){2}{\rule{2.104pt}{0.400pt}}
\multiput(654.00,64.58)(2.808,0.498){93}{\rule{2.333pt}{0.120pt}}
\multiput(654.00,63.17)(263.157,48.000){2}{\rule{1.167pt}{0.400pt}}
\multiput(922.00,112.58)(1.810,0.499){145}{\rule{1.543pt}{0.120pt}}
\multiput(922.00,111.17)(263.797,74.000){2}{\rule{0.772pt}{0.400pt}}
\multiput(1189.00,186.58)(1.351,0.499){195}{\rule{1.179pt}{0.120pt}}
\multiput(1189.00,185.17)(264.553,99.000){2}{\rule{0.589pt}{0.400pt}}
\put(120,31){\usebox{\plotpoint}}
\multiput(120,31)(20.748,0.544){13}{\usebox{\plotpoint}}
\multiput(387,38)(20.650,2.088){13}{\usebox{\plotpoint}}
\multiput(654,65)(20.188,4.821){14}{\usebox{\plotpoint}}
\multiput(922,129)(18.325,9.746){14}{\usebox{\plotpoint}}
\multiput(1189,271)(9.931,18.225){27}{\usebox{\plotpoint}}
\put(1456,761){\usebox{\plotpoint}}
\sbox{\plotpoint}{\rule[-0.400pt]{0.800pt}{0.800pt}}%
\put(120,31){\usebox{\plotpoint}}
\multiput(120.00,32.40)(23.250,0.526){7}{\rule{30.714pt}{0.127pt}}
\multiput(120.00,29.34)(203.251,7.000){2}{\rule{15.357pt}{0.800pt}}
\multiput(387.00,39.41)(5.493,0.504){43}{\rule{8.744pt}{0.121pt}}
\multiput(387.00,36.34)(248.851,25.000){2}{\rule{4.372pt}{0.800pt}}
\multiput(654.00,64.41)(3.317,0.502){75}{\rule{5.429pt}{0.121pt}}
\multiput(654.00,61.34)(256.731,41.000){2}{\rule{2.715pt}{0.800pt}}
\multiput(922.00,105.41)(2.497,0.502){101}{\rule{4.156pt}{0.121pt}}
\multiput(922.00,102.34)(258.375,54.000){2}{\rule{2.078pt}{0.800pt}}
\multiput(1189.00,159.41)(2.245,0.502){113}{\rule{3.760pt}{0.121pt}}
\multiput(1189.00,156.34)(259.196,60.000){2}{\rule{1.880pt}{0.800pt}}
\sbox{\plotpoint}{\rule[-0.800pt]{1.600pt}{1.600pt}}%
\put(120,31){\usebox{\plotpoint}}
\multiput(120,31)(20.748,0.544){13}{\usebox{\plotpoint}}
\multiput(387,38)(20.665,1.935){13}{\usebox{\plotpoint}}
\multiput(654,63)(20.444,3.585){13}{\usebox{\plotpoint}}
\multiput(922,110)(20.113,5.123){14}{\usebox{\plotpoint}}
\multiput(1189,178)(19.923,5.820){13}{\usebox{\plotpoint}}
\put(1456,256){\usebox{\plotpoint}}
\put(776,300){\makebox(0,0)[r]{p=1}}
\put(1176,100){\makebox(0,0)[r]{p=2}}
\end{picture}
\vskip 1.cm
\caption{Occupation numbers $\langle M_p \rangle /2$ as a function of G
for $\Omega = 4$. }
\end{figure}

To obtain a deeper understanding of the above discussed results a study of
the occupation numbers in the different approaches is in order.
They are plotted in Fig. 1 as a function of G. The occupation numbers
$\langle M_p \rangle /2$
obtained using the exact solutions are represented by thin lines, those
obtained using the RPA by small dots, the SCRPA$_1$ by thick lines and
the SCRPA by large dots. Upper curves display the occupations for
p=1, lower curves for p=2.

As it is well known, RPA calculations predict far more correlations than
they are found in the exact ground state. For both states this is evident
here: the small dotted curves increase as function of G faster then the
exact ones, and diverge close to the point of collapse of the RPA. 
On the other hand, we learn from Fig. 1 that the SCRPA under predict the
ground state correlations. The SCRPA$_1$ occupations deviates from the
exact ones nearly as much as the RPA occupations, but underestimating
them. It is remarkable that the SCRPA$_1$ energies are so good, given
these discrepancies.  The SCRPA occupations are very close to the exact
ones.

The correlation functions $\langle M_i M_j \rangle$ are
shown in Table \ref{4.occ} for G=0.4.

\begin{table}
\begin{tabular}{ccccc}
a) &\multicolumn{4}{c}{Exact solutions} \\
   i & j  &  $\langle M_{h _i} M_{h_j} \rangle$ 
&$\langle M_{h _i} M_{p_j} \rangle$ 
&$\langle M_{p _i} M_{p_j} \rangle$ \\ \hline
 1 &1  &0.2556 & 0.1967 & 0.2556 \\
 1 &2  &0.0028 & 0.0616 & 0.0028 \\
 2 &1  &0.0028 & 0.0616 & 0.0028 \\
 2 &2  &0.0864 & 0.0275 & 0.0864
\end{tabular} \\ ~~\\
\begin{tabular}{ccccc}
b) &\multicolumn{4}{c}{SCRPA solutions} \\
   i & j  &  $\langle M_{h _i} M_{h_j} \rangle$ 
&$\langle M_{h _i} M_{p_j} \rangle$ 
&$\langle M_{p _i} M_{p_j} \rangle$ \\ \hline
 1 &1  &0.2251 & 0.1774 & 0.2251 \\
 1 &2  &0.0033 & 0.0603 & 0.0033 \\
 2 &1  &0.0033 & 0.0603 & 0.0033 \\
 2 &2  &0.0816 & 0.0281 & 0.0816
\end{tabular}  \\  ~~\\
\begin{tabular}{ccccc}
c) &\multicolumn{4}{c}{product matrix} \\
   i & j  &  $\langle M_{h _i}\rangle \langle M_{h_j} \rangle$ 
&$\langle M_{h _i}\rangle \langle M_{p_j} \rangle$ 
&$\langle M_{p _i} \rangle \langle M_{p_j} \rangle$ \\ \hline
 1 &1  &0.0127 & 0.0127 & 0.0127 \\
 1 &2  &0.0046 & 0.0046 & 0.0046 \\
 2 &1  &0.0046 & 0.0046 & 0.0046 \\
 2 &2  &0.0017 & 0.0017 & 0.0017
\end{tabular}
\caption{Correlation functions $\langle M_i M_j \rangle$  for
G=0.4, $\Omega = 4$ . 
Insert a) shows the exact results, insert b) the SCRPA results and
insert c) the product of single occupations}
\label{4.occ}
\end{table}

As can be seen in Table \ref{4.occ} the occupation number correlations
obtained
self-consistently from the SCRPA equations are very close to the exact
ones, even for a value of G as large as 0.4. On the other hand the
approximation  $\langle M_i M_j \rangle \approx \langle M_i\rangle \langle 
M_j \rangle $ underestimates the correlations. It could explain the small
occupation numbers obtained when this approximation is
used in the SCRPA equations. 

Although the SCRPA results presented here are in general extremely close
to the exact results, considering correlation or excitation energies,
occupation or correlation numbers, there is a caveat which must be
mentioned here. It concerns the sum rule (\ref{sum}).

\begin{table}
\begin{tabular}{ccc}
&$\sum\limits_{pp'} \langle M_{p} M_{p'} \rangle$
&$ \sum\limits_{hp} \langle M_{h} M_{p} \rangle $
\\ \hline 
exact &  0.3476 & 0.3476 \\
SCRPA & 0.3133 & 0.3260
\end{tabular}
\caption{Sum rule (\ref{sum}) for the exact and SCRPA calculations
for G=0.4, $\Omega = 4$ }
\label{4.sum}
\end{table}

The sums of correlation numbers obtained using the exact calculation and
the SCRPA one are very similar. They are shown in Table
\ref{4.sum}. However, while the exact results fulfill
the sum rule (\ref{sum}), the SCRPA {\em do not} fulfill this sum rule.
This shortcoming certainly stems from the fact that, as we explained in 
Section 8, the SCRPA ground state as defined in Eq. (\ref{vac-rpa}) 
only exists for the special case $\Omega = 2$. In all the other cases
the use of relation (\ref{vac-rpa}) implies some degree of approximation
entailing, for example, the violation of the above sum rule.

\subsection{The $\Omega=10$ case}

The results for $\Omega =10$ repeat the general patterns found for 
$\Omega = 4$. In Table {\ref{10.0} the ground state correlation energies
are
presented as a function of the pairing strength G, in Tables \ref{10.1} 
and \ref{10.2} the
excitation energies of the first and second addition mode in a system with
N=12 fermions, measured relative to the ground state of the system with
N=10 are shown, respectively. 

The SCRPA does not show any collapse, and the energies are far closer to
the exact energies than the RPA ones. For the ground state correlation
energy the ones obtained with the SCRPA$_1$ formalism are slightly closer to
the exact ones than the SCRPA energies, while the opposite is true for the
first and second excited states. For these excitation energies the SCRPA
reproduces with very high accuracy the exact values. It is worth to mention
that the exact excitation energies of the first and second addition modes 
monotonically increase as function of G. This behavior is well reproduced by the
SCRPA, while in the standard RPA these excitation energies are decreasing.
This implies that the screening of the bare interaction is so strong that 
even the sign of the interaction is turned around, bringing the SCRPA solution
closer to the exact one. This is a quite remarkable achievement of SCRPA.

\begin{table}
\begin{tabular}{cccccc}
G  & exact & RPA  & SCRPA$_1$  & SCRPA \\ \hline
.00 &  \,0.0000  &   \,0.0000  &   \,0.0000  &  \,0.0000 \\
.05 &  -0.0086 &  -0.0086  &  -0.0086    &   -0.0086 \\
.10  &  -0.0364 &  -0.0367  &  -0.0365    &   -0.0365 \\
.20  &  -0.1669 &  -0.1756  &  -0.1686    &   -0.1691 \\
.30 &  -0.4350 &  -0.5419  &  -0.4424    &   -0.4497 \\
.33 &  -0.5505 &  -0.8181  &  -0.5594    &   -0.5725 \\
.34 &  -0.5931 &   ***        &  -0.6023    &   -0.6180  \\
.35 &  -0.6379 &   ***        &  -0.6473    &   -0.6658 \\
.36 &  -0.6850 &   ***        &  -0.6943    &   -0.7160
\end{tabular}
\caption{Correlation energies as function of G obtained with exact
calculation, with the RPA, SCRPA$_1$ and SCRPA methods, for $\Omega = 10$.} 
\label{10.0}
\end{table}

\begin{table}
\begin{tabular}{cccccc}
G  & exact & RPA  & SCRPA$_1$  &  SCRPA  \\ \hline
.00 &  1.0000 &  1.0000  &   1.0000   & 1.0000\\
.05 &  1.0003 &  0.9940  &   1.0005   & 1.0003 \\
.10 &  1.0011 &  0.9732  &   1.0034   & 1.0014 \\
.20 &  1.0053 &  0.8604  &   1.0279   & 1.0119 \\
.30 &  1.0143 &  0.5257  &   1.0970   & 1.0539 \\
.33 &  1.0184 &  0.2574  &   1.1266   & 1.0758 \\
.34 &  1.0199 &  ***     &   1.1372   & 1.0840 \\
.35 &  1.0216 &  ***     &   1.1481   & 1.0927 \\
.36 &  1.0233 &  ***     &   1.1592   & 1.1018 
\end{tabular}
\caption{Excitation energy of the first addition mode as function of G
obtained with exact calculation, with the RPA, SCRPA$_1$ and SCRPA
methods, for $\Omega = 10$.} 
\label{10.1}
\end{table}

\begin{table}
\begin{tabular}{cccccc}
G  & exact & RPA  & SCRPA$_1$  &  SCRPA \\ \hline
.00  & 3.0000  & 3.0000   &  3.0000     &  3.0000        \\
.05  & 3.0010  & 2.9971   &  3.0011     &  3.0010 \\
.10  & 3.0056  & 2.9881   &  3.0063     &  3.0057 \\
.20  & 3.0390  & 2.9515   &  3.0460     &  3.0406 \\
.30  & 3.1442  & 2.8935   &  3.1657     &  3.1507 \\
.33  & 3.1998  & 2.8736   &  3.2244     &  3.2070 \\
.34  & 3.2214  & ***      &  3.2467     &  3.2287 \\
.35  & 3.2449  & ***      &  3.2703     &  3.2518 \\
.36  & 3.2701  & ***      &  3.2952     &  3.2765 
\end{tabular}
\caption{Excitation energy of the second addition mode as function of G
obtained with exact calculation, with the RPA, SCRPA$_1$ and SCRPA
methods, for $\Omega = 10$.} 
\label{10.2}
\end{table}

The sum rule (\ref{sum}) is again violated in the $\Omega = 10$ case, as
shown in Table \ref{10.sum}. It must be said that the difference between
both sums, of the order 0.02, must be compared with $N^2 = 100$. The
violation of the sum rule is therefore very small. On the other side, it exhibits
the limitations of the best possible SCRPA treatment of this problem. 

\begin{table}
\begin{tabular}{ccc}
&$\sum\limits_{pp'} \langle M_{p} M_{p'} \rangle$
&$ \sum\limits_{hp} \langle M_{h} M_{p} \rangle $\\ \hline
SCRPA & 0.4509 & 0.4736
\end{tabular}
\caption{Sum rule (\ref{sum}) for the exact and SCRPA calculations for $\Omega = 10$}
\label{10.sum}
\end{table}

\subsection{The $\Omega=24$ and $\Omega=100$ cases}

The correlation energies for $\Omega=24$ are shown in Table \ref{24.0}.
The excitation energies corresponding to the first and second addition
modes are presented in Tables \ref{24.1} and \ref{24.2} respectively.
All the energies are shown as functions of the pairing strength G given in
the first columns. The second column exhibits the exact results in 
Table \ref{24.0}. RPA, r-RPA, SCRPA$_1$ and SCRPA energies are presented in 
the following four columns. 

The large $\Omega$ limit is being approached, where all the RPA descriptions
show their best performances. For the correlation energies even the standard
RPA predictions are good, except for G = 0.25, close to the point of collapse.
For the excitation energies both the RPA and r-RPA results decrease when G
increases, while the SCRPA energies are increasing, as in previous cases, and
in agreement with the exact energies (not shown). At the same time the
differences between the SCRPA$_1$ and SCRPA energies are vanishing. It
strongly supports the use of the simpler SCRPA$_1$ approach for larger $\Omega$,
where the numerical effort needed to solve the SCRPA equations in the fully
self-consistent formalism is not justified.

This is the case for $\Omega = 100$, whose energies are shown in Tables
\ref{100.0}, \ref{100.1} and \ref{100.2} as functions of the pairing strength G. 
Table \ref{100.0} exhibits the exact correlation energies in the second
columns, and the RPA, r-RPA and SCRPA$_1$ in the third, fourth and fifth columns,
respectively. The agreement between the exact and SCRPA is excellent, even for
G = 0.19, where the RPA has collapsed. 

Tables \ref{100.1} and \ref{100.2} present the RPA, r-RPA and SCRPA$_1$ excitation energies 
of the first and second addition modes as function of G. Again only the SCRPA energies
are increasing when G increases, as is expected from the exact results.The latter are
not shown, since it would have needed extra numerical effort without bringing 
new insight to the problem. It should be mentioned that $\Omega = 100$ constitutes
an extremely large configuration space which by no means can be handled with
ordinary diagonalization procedures.

It is well known that the RPA collapse reflects a qualitative change
in the mean field, which after the collapse is dominated by superfluid
correlations \cite{Rin80}. It must be remembered that, 
while the energetics and the occupation
numbers obtained with the SCRPA are very close to the exact ones, the
wave functions around and beyond the value of G at which
standard RPA collapses, being far better than those obtained with RPA
or r-RPA, can nonetheless have an overlap with the exact wave function of
less than 50\% \cite{Hir99}. In this case the SCRPA 
must be extended to the deformed, {\em i.e.} superfluid, basis \cite{Duk90}.

\begin{table}
\begin{tabular}{llllll}
G  & exact & RPA  & r-RPA  & SCRPA$_1$  & SCRPA \\
\hline
0.00  & \,0.0000 & \,0.0000&\,0.0000 &\,0.0000  &\,0.0000   \\
0.05  &  -0.0218 & -0.0218 & -0.0218 & -0.0218 & -0.0218 \\
0.10  &  -0.0953 & -0.0961 & -0.0955 & -0.0955 & -0.0955 \\
0.15  &  -0.2373 & -0.2432 & -0.2358 & -0.2386 & -0.2388 \\
0.20  &  -0.4736 & -0.5096 & -0.4623 & -0.4794 & -0.4809 \\
0.25  &  -0.8452 & -1.1420 & -0.7808 & -0.8607 & -0.8694
\end{tabular}

\caption{Correlation energies as function of G obtained with exact
calculation, with the RPA, r-RPA, SCRPA$_1$ and SCRPA methods for $\Omega = 24$.} 
\label{24.0}
\end{table}

\begin{table}
\begin{tabular}{llllll}
G  &  RPA  & r-RPA  & SCRPA$_1$  & SCRPA \\
\hline
0.00  & 1.0000 & 1.0000 & 1.0000 & 1.0000   \\
0.05  & 0.9911 & 0.9912 & 1.0004 & 1.0001  \\
0.10  & 0.9574 & 0.9594 & 1.0035 & 1.0011  \\
0.15  & 0.8806 & 0.8936 & 1.0153 & 1.0060  \\
0.20  & 0.7155 & 0.7818 & 1.0458 & 1.0233  \\
0.25  & 0.2132 & 0.6279 & 1.1040 & 1.0661  
\end{tabular}

\caption{Excitation energy of the first addition mode as function of G
obtained with the RPA, r-RPA, SCRPA$_1$ and SCRPA methods for $\Omega = 24$.} 
\label{24.1}
\end{table}

\begin{table}
\begin{tabular}{llllll}
G  & RPA  & r-RPA  & SCRPA$_1$  & SCRPA \\
\hline
0.00  & 3.0000 & 3.0000 & 3.0000 & 3.0000    \\
0.05  & 2.9944 & 2.9944 & 3.0007 & 3.0006 \\ 
0.10  & 2.9755 & 2.9761 & 3.0057 & 3.0049 \\ 
0.15  & 2.9398 & 2.9425 & 3.0233 & 3.0200 \\ 
0.20  & 2.8836 & 2.8926 & 3.0705 & 3.0622 \\ 
0.25  & 2.8044 & 2.8297 & 3.1758 & 3.1618 
\end{tabular}

\caption{Excitation energy of the second addition mode as function of G
obtained with the RPA, r-RPA, SCRPA$_1$ and SCRPA methods for $\Omega = 24$.} 
\label{24.2}
\end{table}

\begin{table}
\begin{tabular}{llllll}
G  & exact & RPA  & r-RPA  & SCRPA$_1$  \\
\hline
0.00  & \,0.0000 &\,0.0000 & \,0.0000 & \,0.0000  \\
0.05  &  -0.0946 & -0.0946 & -0.0946 & -0.0946  \\
0.10  &  -0.4250 & -0.4274 & -0.4247 & -0.4256  \\
0.12  &  -0.6467 & -0.6540 & -0.6454 & -0.6483  \\
0.14  &  -0.9365 & -0.9575 & -0.9314 & -0.9405  \\
0.16  &  -1.3127 & -1.3768 & -1.2942 & -1.3217 \\
0.17  &  -1.5413 & -1.6628 & -1.5067 & -1.5543  \\
0.18  &  -1.8027 & -2.0914 & -1.7395 & -1.8209  \\
0.19  &  -2.1028 &   ***   &  ***    & -2.1273
\end{tabular}

\caption{Correlation energies as function of G obtained with exact
calculation, with the RPA, r-RPA and SCRPA$_1$  methods for $\Omega = 100$.} 
\label{100.0}
\end{table}

\begin{table}
\begin{tabular}{llllll}
G  &  RPA  & r-RPA  & SCRPA$_1$  \\
\hline
0.00  & 1.0000 & 1.0000 & 1.0000  \\
0.05  & 0.9857 & 0.9859 & 1.0004  \\
0.10  & 0.9210 & 0.9252 & 1.0056  \\
0.12  & 0.8657 & 0.8774 & 1.0126  \\
0.14  & 0.7762 & 0.8087 & 1.0262  \\
0.16  & 0.6186 & 0.7144 & 1.0504  \\
0.17  & 0.4829 & 0.6575 & 1.0678  \\
0.18  & 0.2181 & 0.5958 & 1.0895  \\
0.19  & ***    &   ***  & 1.1154
\end{tabular}

\caption{Excitation energy of the first addition mode as function of G
obtained with the RPA, r-RPA and SCRPA$_1$ methods for $\Omega = 100$.} 
\label{100.1}
\end{table}

\begin{table}
\begin{tabular}{llllll}
G  & RPA  & r-RPA  & SCRPA$_1$  \\
\hline
0.00  & 3.0000 & 3.0000 & 3.0000    \\
0.05  & 2.9896 & 2.9897 & 3.0006   \\
0.10  & 2.9489 & 2.9499 & 3.0082   \\
0.12  & 2.9193 & 2.9218 & 3.0182   \\
0.14  & 2.8788 & 2.8843 & 3.0381   \\
0.16  & 2.8245 & 2.8362 & 3.0761   \\
0.17  & 2.7912 & 2.8080 & 3.1059   \\
0.18  & 2.7530 & 2.7773 & 3.1457   \\
0.19  &  ***   &  ***   & 3.1985
\end{tabular}

\caption{Excitation energy of the second addition mode as function of G
obtained with the RPA, r-RPA and SCRPA$_1$ methods for $\Omega = 100$.} 
\label{100.2}
\end{table}

\clearpage

\section{Conclusions}

In this work we presented for the first time a fully Self-Consistent
RPA treatment of the picket fence model. This involved developing
equations for some correlation functions which had to be solved 
simultaneously with the SCRPA eigenvalue problem.

One of the remarkable results of the present work is that for 
$\Omega = 2$, {\em i.e.} for the 2 particles case, SCRPA provides
the exact energies of the problem. 
This is not at all trivial because usually many
body approaches deteriorate as the number of particles decreases.
For higher number of levels SCRPA is, of course, not any longer
exact but the results for the correlation energy and the low lying
part of the spectrum are in excellent agreement with the exact ones.

In an earlier work we had used the factorization ansatz 
$\langle N_i N_j \rangle \approx \langle N_i \rangle \langle N_j \rangle $
for the density-density correlation functions. As mentioned
already, we did not make any of these approximations in the present work.
Nevertheless we could verify a posteriori that the above approximation 
yields still excellent results as compared with the exact solutions.
This is more noticeable when $\Omega$ is increased. Actually we found 
that the above factorization approximation also works well for other
models \cite{Duk00}. Therefore one may conjecture that it is quite
generally valid. It is also quite useful, since it reduces the 
numerical work considerably. For example with its use we solved the
SCRPA equations for $\Omega = 100$ without problems.

In spite of the excellent results we could not confirm previous conjectures,
at least for the ansatz used in this work, that SCRPA yields an upper limit
to the ground state energy. Our values are very accurate but slightly
{\bf below} the exact values. This finding is certainly linked to the fact
that for the present case a Self-Consistent RPA ground state wave function
does not exist except for $\Omega = 2$, as we showed in Section 8. This
most likely entails a violation of the Pauli principle which, although small,
in turn gives raise to overbinding. This also explains the slight violation
of the sum rule (67). One should, however, mention that the picket fence model,
because of the low degeneracy of the levels, tests the Pauli principle in a
most sensitive way. This is, for example, born out in the fact that the low lying 
excited states, as a function of G, raise instead of getting lower as one could 
expect from the attractive nature of the interaction. It is another outstanding
result of the present application of the SCRPA theory that its results follow
very closely this trend of the excited states implying that the renormalization
of the bare interaction due to the self-consistency effect can even turn around 
the sign of the interaction.

We also should mention that we have treated only the non-superfluid phase.
It is well known from our previous studies that for interaction strengths 
beyond the collapse of standard RPA one must switch to the `deformed' or
superfluid basis, otherwise the results deteriorate or the iteration
procedure does not converge. The superfluid domain of the present model
shall be studied in a future work.

All in all we can conclude that in this first full application of SCRPA
to a large scale problem the expectation we had gained from earlier studies
in the performance of SCRPA has been fully confirmed. Indeed the results 
did not fail in any major quantitative respect. It may therefore seem 
worthwhile to push the applications of SCRPA to more realistic Hamiltonians.

\section*{Acknowledgments}

This work was supported in part by Conacyt (M\'{e}xico), Conicet (Argentina)
and by the DGES (Spain) under contract BFM2000-1320-C02-02.

\section*{Appendix: Some useful relations}

\ba
\left[ R_\lambda,R^\dagger_{\lambda'}\right] = 
&- \sum_p Y^\lambda_p Y^{\lambda'}_p {\frac {1 - M_p} {1 - \langle M_p
\rangle} }
+ \sum_h X^\lambda_h X^{\lambda'}_h {\frac {1 - M_h} {1 - \langle M_h
\rangle} } ~~,\nonumber\\
\left[ A_\mu,A^\dagger_{\mu'}\right] =
& \sum_p X^\mu_p X^{\mu'}_p {\frac {1 - M_p} {1 - \langle M_p
\rangle} }
- \sum_h Y^\mu_h Y^{\mu'}_h {\frac {1 - M_h } {1 - \langle M_h \rangle } }
~~,\nonumber\\
\left[ A_\mu,R_{\lambda}\right] =   
&- \sum_p X^\mu_p Y^{\lambda}_p {\frac {1 - M_p} {1 - \langle M_p
\rangle} }
+ \sum_h Y^\mu_h X^{\lambda}_h {\frac {1 - M_h} {1 - \langle M_h \rangle }
} ~~,\\
\left[ M_p, R_\lambda \right] = &- 2 Y^\lambda_p 
\{ \sum_{\mu_1} X^{\mu_1}_p A^\dagger_{\mu_1} + 
\sum_{\lambda_1} Y^{\lambda_1}_p R_{\lambda_1} \} \nonumber\\
\left[ M_h, A^\dagger_\mu \right] =  &2 Y^\mu_h 
\{ \sum_{\mu_1} Y^{\mu_1}_p A^\dagger_{\mu_1} + 
\sum_{\lambda_1} X^{\lambda_1}_h R_{\lambda_1} \} \nonumber
\ea

\begin{eqnarray}
\langle R_\lambda R^\dagger_{\lambda_2} R_{\lambda_1} R^\dagger_{\lambda'}
\rangle = \nonumber \\
\sum_{p_1 p_2} {\frac {Y^{\lambda}_{p_1} Y^{\lambda_2}_{p_1}}
{(1 - \langle M_{p_1} \rangle)}}
{\frac {Y^{\lambda_1}_{p_2} Y^{\lambda'}_{p_2}}  
{(1 - \langle M_{p_2} \rangle)}} \langle (1 -M_{p_1}) (1-M_{p_2}) \rangle 
\nonumber \\
-
\sum_{p_1 h_1} {\frac {Y^{\lambda}_{p_1} Y^{\lambda_2}_{p_1}}   
{(1 - \langle M_{p_1} \rangle)}}
{\frac {X^{\lambda_1}_{h_1} X^{\lambda'}_{h_1}}
{(1-\langle M_{h_1} \rangle )}} \langle (1 -M_{p_1}) (1-M_{h_1})
\rangle\\
-
\sum_{p_1 h_1} {\frac {Y^{\lambda_1}_{p_1} Y^{\lambda'}_{p_1}}
{(1 - \langle M_{p_1} \rangle)}}
{\frac {X^{\lambda}_{h_1} X^{\lambda_2}_{h_1}} 
{(1-\langle M_{h_1} \rangle )}} \langle (1 -M_{p_1}) (1-M_{h_1})
\rangle 
\nonumber \\
+
\sum_{h_1 h_2} {\frac {X^{\lambda}_{h_1} X^{\lambda_2}_{h_1}}
{(1-\langle M_{h_1} \rangle )}}
{\frac {X^{\lambda_1}_{h_2} X^{\lambda'}_{h_2}} 
{(1-\langle M_{h_2} \rangle )}} \langle (1 -M_{h_1}) (1-M_{h_2})
\rangle  
\nonumber 
\end{eqnarray}

\begin{eqnarray}
\langle R_\lambda \left[ R_{\lambda_1}, R^\dagger_{\lambda_2} \right]
R^\dagger_{\lambda'} \rangle = \nonumber \\
2 \sum_{p_1 } {\frac {Y^{\lambda_1}_{p_1} Y^{\lambda_2}_{p_1}
Y^{\lambda}_{p_1} Y^{\lambda'}_{p_1}}
{(1 - \langle M_{p_1} \rangle)}} 
-2 \sum_{h_1} {\frac {X^{\lambda_1}_{h_1} X^{\lambda_2}_{h_1}
X^{\lambda}_{h_1} X^{\lambda'}_{h_1}}
{(1 - \langle M_{h_1} \rangle )}}  \nonumber \\
+
\sum_{p_1 p_2} {\frac {Y^{\lambda_1}_{p_1} Y^{\lambda_2}_{p_1}}
{(1 - \langle M_{p_1} \rangle)}}
{\frac {Y^{\lambda}_{p_2} Y^{\lambda'}_{p_2}}  
{(1 - \langle M_{p_2} \rangle)}} \langle (1 -M_{p_1}) (1-M_{p_2}) \rangle 
\nonumber \\
-
\sum_{p_1 h_1} {\frac {Y^{\lambda_1}_{p_1} Y^{\lambda_2}_{p_1}}   
{(1 - \langle M_{p_1} \rangle)}}
{\frac {X^{\lambda}_{h_1} X^{\lambda'}_{h_1}}
{(1 - \langle M_{h_1} \rangle )}} \langle (1 -M_{p_1}) (1-M_{h_1})
\rangle\\
-
\sum_{p_1 h_1} {\frac {Y^{\lambda}_{p_1} Y^{\lambda'}_{p_1}}
{(1 - \langle M_{p_1} \rangle)}}
{\frac {X^{\lambda_1}_{h_1} X^{\lambda_2}_{h_1}} 
{(1- \langle M_{h_1} \rangle )}} \langle (1 -M_{p_1}) (1-M_{h_1}) \rangle 
\nonumber \\
+
\sum_{h_1 h_2} {\frac {X^{\lambda_1}_{h_1} X^{\lambda_2}_{h_1}}
{(1 - \langle M_{h_1} \rangle )}}
{\frac {X^{\lambda}_{h_2} X^{\lambda'}_{h_2}} 
{(1 -\langle M_{h_2} \rangle )}} \langle (1 -M_{h_1}) (1-M_{h_2}) \rangle  
\nonumber 
\end{eqnarray}

\ba
\langle R_\lambda R_{\lambda_1} R^\dagger_{\lambda_2} R^\dagger_{\lambda'}
\rangle =
\langle R_\lambda \left[R_{\lambda_1}, R^\dagger_{\lambda_2}\right]
R^\dagger_{\lambda'} \rangle +
\langle R_\lambda R^\dagger_{\lambda_1} R_{\lambda_2} R^\dagger_{\lambda'}
\rangle \nonumber\\
= 2 \sum_{p_1 } {\frac {Y^{\lambda_1}_{p_1} Y^{\lambda_2}_{p_1}
Y^{\lambda}_{p_1} Y^{\lambda'}_{p_1}}
{(1 - \langle M_{p_1} \rangle)}} 
-2 \sum_{h_1} {\frac {X^{\lambda_1}_{h_1} X^{\lambda_2}_{h_1}
X^{\lambda}_{h_1} X^{\lambda'}_{h_1}}
{(1- \langle M_{h_1} \rangle  )}}  \nonumber \\
+
\sum_{p_1 p_2} 
{\frac  
{ Y^{\lambda_2}_{p_1}  Y^{\lambda'}_{p_2}
 (Y^{\lambda}_{p_1} Y^{\lambda_1}_{p_2} +
 Y^{\lambda_1}_{p_1} Y^{\lambda}_{p_2})}  
{(1 - \langle M_{p_1} \rangle)(1 - \langle M_{p_2} \rangle)}} 
\langle (1 -M_{p_1}) (1-M_{p_2}) \rangle 
\nonumber \\
- 
\sum_{p_1 h_1} {\frac { (Y^{\lambda_2}_{p_1} X^{\lambda'}_{h_1}
+ Y^{\lambda'}_{p_1} X^{\lambda_2}_{h_1})
(Y^{\lambda_1}_{p_1} X^{\lambda}_{h_1} +
 Y^{\lambda}_{p_1} X^{\lambda_1}_{h_1}) }
{(1 - \langle M_{p_1} \rangle)(1- \langle M_{h_1} \rangle )}} 
\langle (1 -M_{p_1}) (1-M_{h_1})
\rangle\\
+
\sum_{h_1 h_2} {\frac { X^{\lambda_2}_{h_1} X^{\lambda'}_{h_2}
( X^{\lambda_1}_{h_1} X^{\lambda}_{h_2} +
  X^{\lambda}_{h_1} X^{\lambda_1}_{h_2}) } 
 {(1- \langle M_{h_1} \rangle )(1- \langle M_{h_2} \rangle )}}
 \langle (1 -M_{h_1}) (1-M_{h_2}) \rangle  
\nonumber 
\ea

\ba
\langle A_\mu A_{\mu_1} A^\dagger_{\mu_2} A^\dagger_{\mu'} \rangle =
 \nonumber\\
= 2 \sum_{h_1 } {\frac {Y^{\mu_1}_{h_1} Y^{\mu_2}_{h_1}
Y^{\mu}_{h_1} Y^{\mu'}_{h_1}} {(1 - \langle M_{h_1} \rangle)}} 
-2 \sum_{p_1} {\frac {X^{\mu_1}_{p_1} X^{\mu_2}_{p_1}
X^{\mu}_{p_1} X^{\mu'}_{p_1}} {(1 - \langle M_{p_1} \rangle )}}  \nonumber
\\
+
\sum_{h_1 h_2} 
{\frac  
{ Y^{\mu_2}_{h_1}  Y^{\mu'}_{h_2}
 (Y^{\mu}_{h_1} Y^{\mu_1}_{h_2} +
 Y^{\mu_1}_{h_1} Y^{\mu}_{h_2})}  
{(1 - \langle M_{h_1} \rangle)(1 - \langle M_{h_2} \rangle)}} 
\langle (1 -M_{h_1}) (1-M_{h_2}) \rangle 
\nonumber \\
- 
\sum_{p_1 h_1} {\frac { (Y^{\mu_2}_{h_1} X^{\mu'}_{p_1}
+ Y^{\mu'}_{h_1} X^{\mu_2}_{p_1})
(Y^{\mu_1}_{h_1} X^{\mu}_{p_1} +
 Y^{\mu}_{h_1} X^{\mu_1}_{p_1}) }
{(1 - \langle M_{p_1} \rangle)(1 - \langle M_{h_1} \rangle )}} 
\langle (1 -M_{p_1}) (1-M_{h_1})
\rangle\\
+
\sum_{p_1 p_2} {\frac { X^{\mu_2}_{p_1} X^{\mu'}_{p_2}
( X^{\mu_1}_{p_1} X^{\mu}_{p_2} +
  X^{\mu}_{p_1} X^{\mu_1}_{p_2}) } 
 {(1 - \langle M_{p_1} \rangle )(1 - \langle M_{p_2} \rangle )}}
 \langle (1 -M_{p_1}) (1-M_{p_2}) \rangle  
\nonumber 
\ea

\end{document}